\documentclass[12pt, utf8]{article}
\usepackage[body={17cm, 23cm, centered}]{geometry}

\usepackage[utf8]{inputenc}
\usepackage[T1]{fontenc}

\usepackage[unicode]{hyperref}
\hypersetup{colorlinks=true,linktocpage,breaklinks,
	urlcolor=blue,
	linkcolor=red,
	citecolor=blue,
    pageanchor=false
}

\usepackage{amsmath,amssymb,amsthm,amscd,cite,comment,amsfonts,indentfirst,color,setspace,bbold}
\usepackage{mathtext,marvosym,textcomp}
\usepackage[makeroom]{cancel}
\usepackage{array}
\usepackage{xcolor}
\usepackage{mathtools}
\usepackage{multirow}
\usepackage{longtable}

\usepackage[vcentermath]{youngtab}

\usepackage{tikz}
\usetikzlibrary{arrows}
\usetikzlibrary{shapes.misc}
\usepackage{tikz-cd}

\tikzset{cross/.style={cross out, draw=black, minimum size=2*(#1-\pgflinewidth), inner sep=0pt, outer sep=0pt},
	cross/.default={5pt}}
\usepackage[ normalem]{ulem}
\usepackage{tocloft}

\numberwithin{equation}{section}

\selectfont

\def\a{\alpha} 
\def\b{\beta} 
\def\g{\gamma} 
\def\d{\delta} 
\def\e{\epsilon}

\def\h{\eta}

\def\l{\lambda} 
\def\m{\mu}
\def\n{\nu} 
 
\def\r{\rho}

\def\s{\sigma} 
  
\def\f{\phi}

\def\Q{\Theta}

\def\L{\Lambda}

\def\W{\Omega}

\def\ba{\bar{a}}

\def\fr{\frac}  \def\dt{\partial}

\def\mc{\mathcal}

\def\mE{\mathcal{E}}
\def\mF{\mathcal{F}}

\def\mL{\mathcal{L}}

\def\mM{\mathcal{M}}
\def\ba{\bar{a}}

\def\XX{\mathbb{X}}

\def\SS{\mathbb{S}}

\def\Ex{\mathrm{E}}
\def\rmSL{\mathrm{SL}}
\def\rmSO{\mathrm{SO}}
\def\rmGL{\mathrm{GL}}
\def\rmUSp{\mathrm{USp}}
\def\rmU{\mathrm{U}}

\def\ba{\begin{array}{r@{}l@{}} }
\def\ea{\end{array}}

\def\baa{\begin{array}{r@{}l  r@{} l@{} } }
\def\ea{\end{array}}

\newcommand{\sfn}{\mathsf{n}}

\newcommand{\AdS}{\mathrm{AdS}}

\newcommand{\rmE}{\mathrm{E}}

\allowdisplaybreaks[4]

\newcommand{\bpm}{\begin{pmatrix}}
\newcommand{\epm}{\end{pmatrix}}

\def\tc{\tilde{c}}
\def\tW{\widetilde{\Omega}}

\newcommand\ol[1]{\overline{#1}}

\begin{document}
\renewcommand{\refname}{\begin{center}References\end{center}}
	
\begin{titlepage}
		
	\vfill
	\begin{flushright}

	\end{flushright}
		
	\vfill
	
	\begin{center}
		\baselineskip=16pt
		{\Large \bf 
		    Polyvector deformations of Type IIB \\backgrounds
		}
		\vskip 1cm
            Kirill Gubarev$^a$\footnote{\texttt{kirill.gubarev@phystech.edu }},
            Edvard T. Musaev$^{a,b}$\footnote{\texttt{musaev.et@phystech.edu}}, 
		    Timophey Petrov$^{a}$\footnote{\texttt {petrov.ta@phystech.edu}},
		\vskip .3cm
		\begin{small}
			{\it 
				$^a$Moscow Institute of Physics and Technology,
			    Institutskii per. 9, Dolgoprudny, 141700, Russia,\\
				$^b$Kazan Federal University, Institute of Physics, Kremlevskaya 16a, Kazan, 420111, Russia
			}
		\end{small}
	\end{center}
		
	\vfill 
	\begin{center} 
		\textbf{Abstract}
	\end{center} 
	\begin{quote}
         We develop a formalism of poly-vector deformations for Type IIB backgrounds with a block diagonal metric and non-vanishing self-dual 5-form RR field strength. Making use of the embedding of the Type IIB theory into the $\mathrm{E}_{6(6)}$ exceptional field theory we derive explicit transformation rules for four-vector deformations. We prove that the algebraic condition following from the Type IIB realization of exceptional Drinfeld algebras is sufficient for the transformation to generate a solution.
	\end{quote} 
	\vfill
	\setcounter{footnote}{0}
\end{titlepage}
	
\clearpage
\setcounter{page}{2}
	
\tableofcontents

\section{Introduction}

Type IIB theory is a phase of M-theory where the perturbative dynamics is well described in terms of closed strings interacting with fields of 10D Type IIB supergravity. These are the metric, the 2-form and the dilaton in the NS-NS sector and $p=0,2,4$ forms in the R-R sector. This field theory also possesses a chiral supersymmetry generated by 32 supercharges collected into the 10D $\mc{N}=(2,0)$ supermultiplet. Therefore, in contrast to Type IIA supergravity this theory cannot be obtained by a simple dimensional reduction of the 11D supergravity that is the low energy description of M-theory\footnote{In fact, the Type IIB theory is a zero-volume limit of the 2-torud reduction of 11D supergravity \cite{Aspinwall:1995fw,Schwarz:1995dk}.}. At the microscopic level this manifests, for example, in the fact D-branes of the Type IIB string theory are not reproduced directly by various wrappings of the M2, M5 membranes and the KK6-monopole. There is however a relation between Type IIA and Type IIB strings stating that when the background has one circular isometric direction these two theories are the same up to trading momentum mode of a string  for its winding mode \cite{Buscher:1987qj,Buscher:1987sk}. When formulated in terms of BPS states of the theory this T-duality symmetry together with the S-duality symmetry of Type IIB can uplifted to the full U-duality symmetry of M-theory \cite{Hull:1994ys,Obers:1998fb}. Hence, although Type IIB degrees of freedom are not related to that of M-theory by a simple dimensional reduction one is able to relate them by making use of U-duality symmetry. This fact stands behind the construction of exceptional field theory that unifies Type IIA/B and 11D supergravity theory in a single field theoretical framework covariant under U-duality symmetries. The latter are realised as coordinate transformations in a special extended space (see \cite{Hohm:2013pua,Hohm:2013vpa,Musaev:2014lna} for some of the original papers and \cite{Hohm:2019bba,Musaev:2019zcr,Berman:2020tqn} for a recent review).

Recently in a series of works \cite{Bakhmatov:2019dow,Bakhmatov:2020kul,Gubarev:2020ydf,Musaev:2023own} a solution generation technique for 11D supergravity has been developed that is heavily based on symmetries exceptional field theory. To be more concrete a solution generating transformation is parametrised by a polyvector made of Killing vectors $k_a{}^m$ of the initial background and belongs to a U-duality group (SL(5) or $\Ex_{6(6)}$ in the referred works, however any other duality group will also do). The polyvector has the following general form 
\begin{eqnarray}
    \W^{m_1\dots m_p} = \fr1{p!}\r^{a_1\dots a_p}k_{a_1}{}^{m_1}\dots k_{a_p}{}^{m_p},
\end{eqnarray}
where $a$ label Killing vectors and $m_i$ stand for tensor component indices. The condition that the transformed background is a solution to 11-dimensional supergravity equations are the so-called generalized Yang-Baxter equation and unimodularity condition. The former has been first obtained in the context of exceptional Drinfeld algebras in \cite{Sakatani:2019zrs,Malek:2019xrf} and is a natural generalisation of the classical Yang-Baxter equation (CYBE) appearing in classical integrable systems. It is worth to note, however, that its quantum formulation is not known. An attempt to suggest one has been made in \cite{Malek:2020hpo}, a review on possible relations between generalized CYBE and 3d integrability can be found in \cite{Gubarev:2023jtp}. Based on to the labelling of generators of a U-duality group in terms of its largest geometric $\rmGL(n)$ subgroup one concludes that polyvector deformations of 11D backgrounds are parametrised by a 3-vector and a 6-vector (polyvectors of mixed symmetry might also contribute for $\rmE_{7(7)}$ and higher duality groups). Polyvector deformations of Type IIA backgrounds are trivially reproduced by considering 11D backgrounds with a circular isometric direction and reside in the standard bi-vector Yang-Baxter deformations and additional 1-, 3-, 5- and 6-vector deformations.

In this paper we construct a framework for polyvector deformations of  10-dimensional Type IIB backgrounds by considering embedding of the Type IIB theory into the $\Ex_{6(6)}$ exceptional field theory (provided in details e.g. in \cite{Baguet:2015xha}). For simplicity we restrict the narrative to backgrounds of the form $M_5\times N_5$, meaning that the $N_5$ block of the metric does not depend on ``external'' coordinates of the manifold $M_5$, and the external block in tern depends on the ``internal'' coordinates only by an overall prefactor. We also suppose that no non-diagonal elements present in the metric and the the only non-trivial gauge field contribution is given by the 5-form field strength $F_5$. In particular, this set-up covers the AdS$_5\times \SS^5$ background, dual to the $\mc{N}=4$ $D=4$ super Yang-Mills (SYM) theory, and describes its exactly marginal deformations when the Killing vectors are taken along the 5-sphere. As it has been shown in \cite{Musaev:2023own} in contrast to bi-vector deformations one is able to perform polyvector transformation along non-abelian compact isometries and hence to generate a larger class of exactly marginal deformations. A general discussion of the relation between (bi-vector) transformations of supergravity backgrounds and deformations of the dual field theory can be found in \cite{Imeroni:2008cr} and also in the work \cite{Lunin:2005jy} where an example has been provided for the Leigh-Strassler $\beta$-deformation \cite{Leigh:1995ep}.  We present an example of a 4-vector deformation of AdS$_5\times \SS^5$ that is a wedge product of a 3-vector of \cite{Musaev:2023own} and an abelian generator of the $\rmSO(6)$ symmetry group. As expected bi-vector deformations coming from the NS-NS and the R-R sector appear combined into an SL(2) doubled, each component of which must satisfy CYBE (in the absense of 4-vectors), and hence is subject to the theorem of \cite{Lichnerowicz:1988abc,Pop:2007abc} forbidding deformations on non-abelian compact isometries.

The paper is structured as follows. In Section \ref{sec:embed} we briefly remind the construction that embeds the Type IIB supergravity into the $\rmE_{6(6)}$ exceptional field theory, derive full generalized metric by exponentiating the corresponding generators and describe the truncation. In Section \ref{sec:deform} we provide explicit form of the 2- and 4-vector deformation matrices and derive explicit expressions for transformations of Type IIB supergravity fields under a 4-vector deformation. We show that given an algebraic condition such deformations in the poly-Killing ansatz map solutions into solutions. In Section \ref{sec:example} we illustrate the technique by the example of a 4-vector deformation of the $\AdS_5\times \SS^5$ background. Finally, Section \ref{sec:conclusions} summarizes the paper and Appendicies provide information on our notations and contain some technical details of calculations of the main text.

\section{Embedding of Type IIB into the \texorpdfstring{$E_6$}{E6} ExFT}
\label{sec:embed}

Following the general idea of \cite{Bakhmatov:2019dow,Bakhmatov:2020kul} polyvector deformations of a supergravity background (be it 11D or Type IIB) are most straightforwardly constructed using the exceptional field theory parametrisation of supergravity fields. Since such a deformation is a particular (local) U-duality transformation it is realised as a linear group action in this parametrization. The relation between fields of ExFT, that transform linearly, and the supergravity fields in the standard formulation, that transform non-linearly, is far from obvious and basically originates from the dualization rules of \cite{Cremmer:1997ct}. A given exceptional field theory, say with the $\rmE_{6(6)}$ group relevant here, contains both 11D supergravity and Type IIB supergravity depending on how the exceptional group is broken to its so-called geometric subgroup $\rmGL(n)$, where $n=6$ for the 11D embedding and $n=5$ for the Type IIB embedding. In the latter case the S-duality $\rmSL(2)$ subgroup survives the projection. Below we will follow the logic that allows to define polyvector deformations for Type IIB backgrounds in the most straightforward way. First we define their action on the fields of the $\rmE_{6(6)}$ exceptional field theory. This is certainly different from that of \cite{Gubarev:2020ydf} since Killing vectors, that define a deformation, belong to the fundamental irrep of a different  geometric $\rmGL(n)$ subgroup of the U-duality group. Second, by making use of the known relation between Type IIB fields in the $10=5+5$ split and the $\rmE_{6(6)}$ ExFT we arrive at explicit transformation rules for supergravity fields and present an example. In this Section

\subsection{Field content and geometric embedding}

Exceptional field theory is a way to write both 11D and Type IIB supergravity as a field theory generally covariant under local U-duality symmetries. Here we are interested in the $\rmE_{6(6)}$ theory where all fields are collected in irreps of $\rmE_{6(6)}$ and local diffeomorphisms include transformations in the corresponding algebra and act on coordinates of a special extended space. Let us briefly highlight relevant for us aspects of this theory and its relation to the Type IIB supergravity referring the reader for more details to the original works \cite{Hohm:2013pua,Hohm:2013vpa} and review papers \cite{Hohm:2019bba,Musaev:2019zcr,Berman:2020tqn}.

The $\rmE_{6(6)}$ exceptional field theory is defined on a space-time parameterized by 5 coordinates $x^\m$ (the so-called external) and 27 coordinates $Y^M$ (the so-called internal). Dependence of all fields of the theory on the latter is subject to a constraint called section condition\footnote{The notion of the section constraint was first introduced in \cite{Siegel:1993th} in the context of string field theory and was a coordinate manifestation of the level matching condition. A general treating of the constraint as the consistency condition for the algebra of generalized Lie derivatives was presented in \cite{Berman:2012vc}.}
\begin{equation}
    d^{MNK}\dt_M \bullet \dt_N \bullet = 0,
\end{equation}
where $d^{MNK} = d^{(MNK)}$ is the fully symmetric invariant tensor of $\rmE_{6(6)}$ and bullets stand for any field of the theory or a combination of fields and their derivatives. This condition is a consistency constraint for the algebra of generalized Lie derivatives to which we will return in a moment. 

The section constraint can be solved in two different ways by allowing all the fields to depend only on a subset of 6 or 5 coordinates out of 27. This corresponds to embedding of the maximal $\rmGL(6)$ and $\rmGL(5)\times \rmSL(2)$ subgroups respectively. Restricting coordinate dependence according to the first embedding gives the standard 11D supergravity. The latter gives the Type IIB embedding and the fundamental $\bf 27$ of the $\rmE_6$ decomposes as follows 
\begin{align}
{\bf \ol{27} } \rightarrow {\bf (5,1)}_{-4}+{\bf (\ol{5},2)}_{-1}+{\bf (10,1)}_2+{\bf (1,2)}_{5}.
\end{align}
where the first and the second numbers in brackets denote irreps of the $\rmSL(5)$ and $\rmSL(2)$ respectively, and the subscript denotes level w.r.t. to the $\rmGL(1)$ group. In terms of tensor components the decomposition reads 
\begin{equation}
    \ \left\{Y^M\right\} \rightarrow\left\{ \,y^m\,,\; y_{m\,\alpha}\,,\; y^{mn}\,,\;  y_\alpha \,\right\}
\;.
\label{cord_split}\end{equation}
Here small Latin indices from the middle of the alphabet $m,n,k,\dots = 1,\dots,5$ label the \textbf{5} of $\rmGL(5)$, $y^{m n}=y^{[m n]}$ transforms in the \textbf{10}, small Greek indices $\alpha,\beta,\dots = 1.,2$ label the fundamental representation of $\rmSL(2)$. All index conventions used in this text are collected in Appendix \ref{app:index}.
In what follows we will always assume that all fields are functions of only $y^m$, that is we are in the Type IIB sector of the $\rmE_{6(6)}$ ExFT. However, for convenience and to keep expressions covariant we will keep using $\dt_M$ in some expressions.

The bosonic field content of $\rmE_{6(6)}$ ExFT is represented by the following fields (for its supersymmetric extension see \cite{Musaev:2014lna})
\begin{equation}\label{E6fields}
    \begin{aligned}
     \{ & g_{\m\n}, && \mM_{MN}, && A_\m\,^{M}, && B_{\m\n M}\},
    \end{aligned}
\end{equation}
where $g_{\m\n}$ is  the metric of the external space, $\mM_{MN}$ is the so-called generalized metric parametrising the scalar coset such that $\mM_{MN} \in {\rmE_{6(6)}}/{\rmUSp(8)}$, $A_\m\,^{M}$ is a generalized connection and $B_{\m\n M}$ is a set of two-forms. In what follows we will truncate this set by keeping $A_\m\,^{M}=0$, and $ B_{\m\n M}=0$. Hence our formalism in the presented shape is applicable only to Type IIB backgrounds of a certain form. Certainly, it can be extended pretty much as in \cite{Barakin:2024abc}.

As it has been shown in \cite{Gubarev:2020ydf} such a truncation of the theory allows to write it completely in terms of scalar generalized fluxes, which are basically coordinate dependent components of the embedding tensor of the maximal $D=5$ gauged supergravity. These are defined in the following way 
\begin{equation}\label{LieE2}
    \mL_{E_{A}} E^{M}{}_{B} = \mF_{A,B}{}^{C} E^{M}{}_{C},
\end{equation}
with the components reading explicitly
\begin{equation}\label{Fstructure}
    \mF_{A,B}{}^{C} = 2 E_{M}{}^{C} E^{K}{}_{[A|} \dt_K E^{M}{}_{|B]} + 10 d^{MKR} d_{NLR} E_{M}{}^{C} E^{N}{}_{B} \dt_K E^{L}{}_{A}.
\end{equation}
We denote $\mL$ the so-called generalized Lie derivative defined as 
\begin{equation}
   \mL_{\Lambda} V^M=   \Lambda^K \partial_K V^M-\partial_K\Lambda^M V^K
  +\Big(\lambda-\frac{1}{3}\Big)\,\partial_P\Lambda^P\, V^M
  +10\, d_{NLR}\, d^{MKR}  \partial_K\Lambda^L V^N,
\label{gen_lie}
\end{equation}
for an arbitrary vector $V^M$ of weight $\l$. Since the transformation parameter $\L^M$ must have weight 1/3, to define fluxes as in \eqref{LieE2} a rescaled generalized vielbein $E_M{}^A$ is defined to have the proper weight $\l = 1/3$. Its generalized metric 
\begin{equation}\label{metricvielbein}
    M_{MN} = E_M{}^A E_N{}^B M_{AB}.
\end{equation}
is related to $\mc{M}_{MN}$ parametrizing the scalar coset by 
\begin{equation}
\label{rescalings}
    \mM_{MN}  = e_{(5)}^{\fr25} M_{MN},
\end{equation}
where $e_{(5)} = \det e_\m{}^a$ is determinant of the 5-dimensional external vielbein. Similarly for the  generalized metric $\mc{M}_{MN}$ of zero weight we have
\begin{equation}\label{metricvielbein0}
    \mM_{MN} = \mE_M{}^A \mE_N{}^B \mM_{AB}.
\end{equation}
Since field equations of the truncated $\rmE_6$ exceptional field theory can be written completely in terms of $\mF_{A,B}{}^C$, to obtain a new solution by a transformation of a given solution it is sufficient to require invariance of $\mF_{A,B}{}^C$. Such an approach has first been advocated in \cite{Borsato:2020bqo} for bivector deformations and further generalized to polyvector deformations in \cite{Gubarev:2020ydf}. 

The crucial observation now is the following: the above obviously statement holds for any reduction of ExFT, be it to the 11D or Type IIB theories. Hence, it is sufficient to check algebraic conditions on polyvector deformation parameters (that will certainly be different of that in the 11D case) for the corresponding transformation to keep the fluxes invariant. To write such a transformation explicitly in terms of Type IIB fields and to apply to specific backgrounds we will need to do some work on the mapping between Type IIB fields in the standard KK decomposition and ExFT field multiplets.

\subsection{Truncated ExFT in IIB parametrization}

\subsubsection{Generators and the generalized metric}

Under the action of the geometric $\rmGL(5)$ subgroup together with the S-duality group $\rmSL(2)$ the fundamental $\bf \ol{27}$ and the adjoint irreps of $\rmE_{6(6)}$ are decomposed as follows
\begin{equation}
\begin{aligned}
{\bf \ol{27} } &\rightarrow {\bf (5,1)}_{4}+{\bf (\ol{5},2)}_{1}+{\bf (10,1)}_{-2}+{\bf (1,2)}_{-5} ,\\
{\bf 78 } & \rightarrow  {\bf (5,1)}_{-6} +{\bf (\ol{10},2)}_{-3} + {\bf (1,3)}_{0}+{\bf (24 + 1,1)}_{0}+ {\bf (10,2)}_{3}+{\bf (\ol{5},1)}_{6}.
\label{split27B}
\end{aligned}
\end{equation} 
In terms of tensor components the first line is written as in \eqref{cord_split}, while the second line reads
\begin{equation}
    \{ T_{\bf \mathrm{A}}  \} = \{T_{m_1\dots m_4}, T_{mn}{}^\a, T_{(\a\b)}, T_m{}^n, T_\a{}^{mn},T^{m_1\dots m_4}\}
\end{equation}
In these notations the standard prescription for non-linear realisation is that the generalized metric is generated by generators of non-negative level (see e.g. \cite{Gubarev:2020ydf}). Hence for the generalized vielbein of weight zero we write
\begin{equation}
    \mE=\exp\left({T_\gamma{}^{mn} B_{mn}{}^{\gamma}}\right)\exp\left(T^{m_1 \ldots m_4} c_{m_1 \ldots m_4}\right) \mc{V}_{\rmGL(5)}\mc{V}_{\rmSL(2)\times \rmGL(1)},
\label{gen_reper}
\end{equation}
where the fields $B_{mn}{}^\a$ and $c_{mnkl}$ will be related to the doublet constructed from the NS-NS and R-R 2-forms and to the RR 4-form gauge potential, and we define
\begin{equation}
    \begin{aligned}
        \mc{V}_{\rmGL(5)} & = e^{\frac{1}{3}}||e_{m}{}^{a}||,\\
        \mc{V}_{\rmSL(2)\times \rmGL(1)} & = e^{-\frac{2}{3}} ||\nu_{\alpha}{}^{\overline{\alpha}}||.
    \end{aligned}
\end{equation}
The matrix $e_m{}^a$ is the standard vielbein defining a metric $g_{m n}=e_{m}{}^{a} e_{n}{}^{b} \h_{a b}$ in the 5-dimensional space parameterised by $y^m$. The matrix $m_{\alpha \beta}=\nu_{{\alpha}}{}^{\overline{\alpha}} \nu_{{\beta}}{}^{\overline{\beta}} \delta_{\overline{\alpha} \overline{\beta}} $ encodes the axio-dilaton of the Type IIB theory, and we also include a factor of $e = \det e_m{}^a$ into its definition for further convenience.

The invariant totally symmetric $d$-tensor of $\rmE_6$ can be written in terms of the invariant tensors of $\rmSL(5)$ and $\rmSL(2)$ as follows \cite{Hohm:2013vpa,Baguet:2015xha} 
\begin{equation}
\begin{aligned}
 & d^{MNK} : &&
 d^{m}{}_{n\alpha,\beta} = \frac1{\sqrt{10}}\, \delta_n{}^m \epsilon_{\alpha\beta}\;,&& 
d^{mn}{}_{k\alpha,l\beta} = \frac1{\sqrt{5}}\, \delta^{mn}_{kl}\,\epsilon_{\alpha\beta}\;, &&
d^{mn,kl,p} = \frac1{\sqrt{40}}\,\epsilon^{mnklp}\;, \\
& d_{MNK} :  && d_{m}{}^{n\alpha,\beta} = \frac1{\sqrt{10}}\, \delta_m{}^n \epsilon^{\alpha\beta}\;, &&
d_{mn}{}^{k\alpha,l\beta} = \frac1{\sqrt{5}}\, \delta_{mn}^{kl}\,\epsilon^{\alpha\beta}\;, &&
d_{mn,kl,p} = \frac1{\sqrt{40}}\,\epsilon_{mnklp}\;.
\end{aligned}
\end{equation}
Explicitly invariance of the $d$-tensor in the fundamental representation can be expressed as the following condition
\begin{equation}
    T_{(M}{}^{N}d_{|N|KL)}=0.
    \label{d_tensor_cond}
\end{equation}
The matrix representation of the generators can be found in Appendix \ref{app:E6gens}.

The generalized metric of weight zero  ${\cal M}_{MN}\in{\rmE_{6(6)}/\rmUSp(8)}$ is defined as usual as  
\begin{align}
    &{\cal M}_{MN}=\mE_M{}^{A}\mE_N{}^{B} \delta_{A B}.
\label{gen_mat_from_rep}\end{align}
To fix the notations let us provide explicit form of the flat metric $\d_{AB}$ 
\begin{equation}
\delta_{A B}=\begin{pmatrix}
 \pm \eta_{ a b} & 0 & 0 & 0 \\
 0 &  \delta^{\overline{\alpha}\overline{\beta}}\, \eta^{a b}  & 0 & 0 \\
 0 & 0 & \, \eta_{a_1 [a_2} \eta_{| b_1| b_2]} & 0 \\
 0 & 0 & 0 &  \pm \delta^{\overline{\alpha}\overline{\beta}}\, 
\end{pmatrix}.
\end{equation}
where  $\h_{a b}=\mathrm{diag}\left(\pm 1,1,1,1,1\right)
$ is the flat $d=5$ metric. Here and everywhere below the upper sign corresponds to the case when the extended space does not include the time direction and the lower sign --- when it does. Note that antisymmetrization of indices has a factor of $1/2$. Using the definition of the generalized vielbein \eqref{gen_reper} and the explicit of of the generators in Appendix \ref{app:E6gens} one is able to derive explicitly components of the generalized metric in terms of the $d=5$ tensor fields (related to the Type IIB supergravity fields). Most convenient is to  start with labelling all components of the generalized metric ${\cal M}_{MN}$ as follows
\begin{equation}
\label{M2B}
{\cal M}_{KM} = 
\begin{pmatrix}
{\cal M}_{km}&{\cal M}_k{}^{m\beta}&{\cal M}_{k,mn}&{\cal M}_{k}{}^{\beta}\\
{\cal M}^{k\alpha}{}_{m}&{\cal M}^{k\alpha,}{}^{m\beta}&{\cal M}^{k\alpha}{}_{mn}&{\cal M}^{k\alpha,}{}^{\beta}\\
{\cal M}_{kl,m}&{\cal M}_{kl}{}^{m\beta}&{\cal M}_{kl,mn}&{\cal M}_{kl}{}^{\beta}\\
{\cal M}^\alpha{}_{m}&{\cal M}^{\alpha,m\beta}&{\cal M}^\alpha{}_{mn}&{\cal M}^\alpha{}^{\beta}
\end{pmatrix}\;.
\end{equation} 
where explicit expression for each block can be found using \eqref{gen_reper} and \eqref{gen_mat_from_rep}:
\begin{equation}
\begin{aligned}
    \mM_{m n}&=\pm \left| g \right|^{\frac{1}{3}}g_{m n} -  4\left| g \right|^{\frac{1}{3}}\epsilon_{\beta' \beta}B^\beta_{mn'}m^{\beta' \alpha'}g^{n' m'}\epsilon_{\alpha ' \gamma'}B^{\gamma '}_{m' n} + \\&+ \left| g \right|^{-\frac{2}{3}} T^{n_1 n_2}_{m}  g_{n_1 n_2, \, m_1' m_2'}\widetilde{T}^{m1' m2'}_{n} \pm \left| g \right|^{-\frac{2}{3}} R^{n_1 n_2 }_{ m\beta \alpha} B^{\alpha}_{n_1 n_2}  m^{\beta \alpha'}\widetilde{R}^{n_1 n_2 }_{ n\alpha' \l} B^{\l}_{n_1 n_2}\\
    \mM^{m \alpha}_{ n}&=  2\left| g \right|^{\frac{1}{3}}m^{\alpha \alpha'} \epsilon_{\alpha' \gamma'} g^{m m'}B^{\gamma '}_{m' n}+ \\ 
    &\quad + \left| g \right|^{-\frac{2}{3}} \sqrt{2}\epsilon^{\beta \alpha}\,\widetilde{B}_\beta^{m n_1n_2}g_{n_1 n_2, \, m_1' m_2'}\widetilde{T}^{m1' m2'}_{n} \pm \left| g \right|^{-\frac{2}{3}} W^{m \alpha}_{\beta} m^{\beta \alpha'}\widetilde{R}^{n_1 n_2 }_{ n\alpha' \l} B^{\l}_{n_1 n_2}\\
    \mM_{m1 m2, n}&=  \left| g \right|^{-\frac{2}{3}}g_{m_1 m_2, \, m_1' m_2'}\widetilde{T}^{m1' m2'}_{n} \pm \left| g \right|^{-\frac{2}{3}} \sqrt{2}\, \epsilon_{\beta\gamma}B_{m_1m_2}^\gamma m^{\beta \alpha'}\widetilde{R}^{n_1 n_2 }_{ n\alpha' \l} B^{\l}_{n_1 n_2}\\
    \mM^{\alpha}{}_{ n}&= \pm \left| g \right|^{-\frac{2}{3}}m^{\alpha \alpha'}\widetilde{R}^{n_1 n_2 }_{ n\alpha' \beta} B^{\beta}_{n_1 n_2}\\
    \mM^{\alpha m,\beta n}&=    \left| g \right|^{\frac{1}{3}}m^{\alpha \beta} g^{m n}+ \\
    &\quad + 2\left| g \right|^{-\frac{2}{3}} \,\epsilon^{\gamma' \alpha}\widetilde{B}_{\gamma'}^{m n_1n_2}g_{n_1 n_2, \, m_1' m_2'}\,\epsilon^{\gamma \beta}\widetilde{B}_{\gamma}^{m_1 'm_2 ' n }  \pm  \left| g \right|^{-\frac{2}{3}} W^{m \alpha}_{\beta'} m^{\beta' \alpha'}W^{n \beta}_{\alpha'}\\
    \mM^{n \beta}_{m1 m2}&=   \sqrt{2}\left| g \right|^{-\frac{2}{3}}g_{m_1 m_2, \, m_1' m_2'}\,\epsilon^{\gamma \beta}\widetilde{B}_{\gamma}^{m_1 'm_2 ' n } \pm \sqrt{2}\left| g \right|^{-\frac{2}{3}} \, \epsilon_{\beta' \gamma}B_{m_1m_2}^{\gamma} m^{\beta' \alpha'} W^{n \beta}_{\alpha'}\\
    \mM^{\alpha,n \beta}&=  \pm \left| g \right|^{-\frac{2}{3}}m^{\alpha \alpha'}W^{n \beta}_{\alpha'}\\
 \mM_{m_1 m_2, n_1 n_2}&=\left| g \right|^{-\frac{2}{3}}g_{m_1 m_2, \, n_1 n_2} \pm \left| g \right|^{-\frac{2}{3}} 2\, \epsilon_{\beta \l}B_{m_1m_2}^\l m^{\beta \alpha'} \, \epsilon_{\alpha ' \gamma}B_{n_1 n_2 }^{\gamma} \\
\mM^{\alpha}_{ n_1 n_2}&=\pm \left| g \right|^{-\frac{2}{3}}m^{\alpha \gamma} \epsilon_{\gamma \alpha'}\sqrt{2}\, B_{n_1 n_2 }^{\alpha '} \\
 \mM^{\alpha \beta}&=
\pm\left| g \right|^{-\frac{2}{3}}m^{\alpha \beta} 
\end{aligned}
\label{our_generalized_metric_0}
\end{equation}

\begin{equation}
\label{our_generalized_metric_non_renormS}
\end{equation}

where were used the following definitions:
\begin{equation}
\begin{aligned}
 g_{n_1 n_2, m_1 m_2}&= g_{n_1 [m_1 }g_{|n_2| m_2]},\\
 B_{mn \a} & = \e_{\a\b} B_{mn}{}^\b, &&  & T^{n1 n2}_{m}&=\sqrt{2}\left|g\right|^{\frac{1}{2}} \,B^\gamma_{m  p}\,\widetilde{B}_\gamma^{ p  n_1 n_2}+\frac{1}{\sqrt{2}} \left|g\right|^{\frac{1}{2}} \tc_m^{n_1 n_2}\, ,\\
 B_{mn}{}^{\a} & =  B_{mn \b}\e^{\b\a} , && & R^{n_1 n_2 }_{m \beta \alpha}&=-\frac{2}{3}\,\epsilon_{\alpha \gamma} \left|g\right|^{\frac{1}{2}} B^\gamma_{m n}\,\widetilde{B}_\beta^{ n  n_1 n_2}+ \epsilon_{ \beta \alpha} \left|g\right|^{\frac{1}{2}}\tc_{m}^{n_1 n_2}\,, \\
 \widetilde{B}_\alpha^{m n_1 n_2} & = \frac{1}{2}\,\epsilon_{\alpha\gamma}\,\varepsilon^{m  n_1 n_2  p_1 p_2}\,B^\gamma_{ p_1 p_2}\,, &&  & W^{m \alpha}_{\beta}&=\left|g\right|^{\frac{1}{2}}\,\widetilde{B}_\beta^{m p_1 p_2}\,B^\alpha_{p_1 p_2}- \delta_{\beta}^{\alpha}  \left|g\right|^{\frac{1}{2}} \tc^{m}, \\ 
 \tc_m^{ n_1 n_2} & \equiv \frac{1}{3!}\,\varepsilon^{ n_1 n_2  p_1 p_2 p_3}\,c_{m  p_1 p_2 p_3}\,, \\
 \tc^m &= \frac{1}{4!}\,\varepsilon^{m  p_1\cdots  p_4}\,c_{ p_1\cdots  p_4}\,.
\end{aligned}
\end{equation}
The generalized vielbein that defines fluxes and the corresponding metric are related to the generalized vielbein and the metric of weight zero by the rescaling \eqref{rescalings}. In what follows we will consider a class of backgrounds were the external (the space-time) part of the metric depends on the internal coordinates only by a prefactor. This is the case, for example, for the $\AdS_5 \times \SS^5$ background. For this reason it is more convenient to define the rescaling more specifically as
\begin{equation}
\label{renorm_formulas}
\begin{aligned}
    M_{M N}&=e^{2\phi}\left|g\right|^{-\frac{1}{3}}\mM_{M N},\\
    g_{\m\n}=g_{\m\n}(x^\m, y^m)&=e^{-2\phi\left(y\right)}\left|g\right|^{\frac{1}{3}}\bar{g}_{\mu \nu}\left(x\right),
\end{aligned}
\end{equation}
where $\phi$ is constant along Killing vectors of supergravity metric, that came from KK ansatz, i.e. it's weight in \eqref{gen_lie} equal to zero.
The corresponding rescaled generalized vielbein, defined as usual as
\begin{equation}
    M_{M N}=E_{M}{}^{A}E_{N}{}^{B}\delta_{A B}
\end{equation}
will be used to define generalized fluxes.

\subsubsection{Type IIB fields}

In our setup the only non-vanishing fields of exceptional field theory are the external metric $g_{\m\n}$ and the generalized metric $M_{MN}$. Under an $E_{6(6)}$ transformation these transform linearly and do not generate other fields. Hence, the only $\rmGL(5)$ tensor fields we will be dealing with are
\begin{equation}
    \begin{aligned}
        &g_{\m\n}, && g_{mn}, && c_{mnkl}, && B_{mn \a}.
    \end{aligned}
\end{equation}
To proceed with deformations of particular solutions to Type IIB field equations it is then necessary to map these to (components of) the Type IIB supergravity fields. The latter include the metric, dilaton, the RR scalar, the doublet of 2-forms and the 4-form RR gauge field with self-dual field strength:
 \begin{equation}
  G_{\hat{\mu} \hat{\nu}}\;, \quad 
  m_{\alpha\beta}\;,  \quad
  \hat{C}_{\hat{\mu}\hat{\nu}}{}^{\alpha}\;, \quad
  \hat{C}_{\hat{\mu}\hat{\nu}\hat{\rho}\hat{\sigma}}\;, 
  \label{IIBfields}
\end{equation}
where $\alpha,\beta=1,2\;$, $\hat{\mu},\hat{\nu},\ldots \in 0..9$ are the ten--dimensional curved 
indices. The dilaton and the RR scalar a combined into the matrix $m_{\a\b} \in {\rmSL}(2)/{\rmSO}(2)$. The field strengths for the gauge potentials are defined as
\begin{equation}
    \begin{aligned}
  \hat{F}_{\hat{\mu}\hat{\nu}\hat{\rho}}{}^{\alpha} & =  
  3\,\partial_{[\hat{\mu}}\hat{C}_{\hat{\nu}\hat{\rho}]}{}^{\alpha}\;. \\
   \hat{F}_{\hat{\mu}_1\ldots\hat{\mu}_5} &= 5\,\partial_{[\hat{\mu}_1}\hat{C}_{\hat{\mu}_2\ldots \hat{\mu}_5]}
  -\frac54 \, 
  \varepsilon_{\alpha\beta}\,\hat{C}_{[\hat{\mu}_1\hat{\mu}_2}{}^{\alpha}\hat{F}_{\hat{\mu}_3\hat{\mu}_4\hat{\mu}_5]}{}^{\beta}\;, 
 \label{F5}
    \end{aligned}
\end{equation} 
and the latter satisfies the following self-duality equations
\begin{equation}
\hat{F}_{\hat{\mu}\hat{\nu}\hat{\rho}\hat\sigma\hat\tau}=
\frac1{5!}\,\sqrt{|{\rm det}\,G_{\hat\mu\hat\nu}|}\,\epsilon_{\hat{\mu}\hat{\nu}\hat{\rho}\hat\sigma\hat\tau
\hat\mu_1\hat\mu_2\hat\mu_3\hat\mu_4\hat\mu_5}\,
\hat{F}^{\hat\mu_1\hat\mu_2\hat\mu_3\hat\mu_4\hat\mu_5}
\;,
\label{FF55}
\end{equation}
with $\epsilon_{0,...,9}=1$.

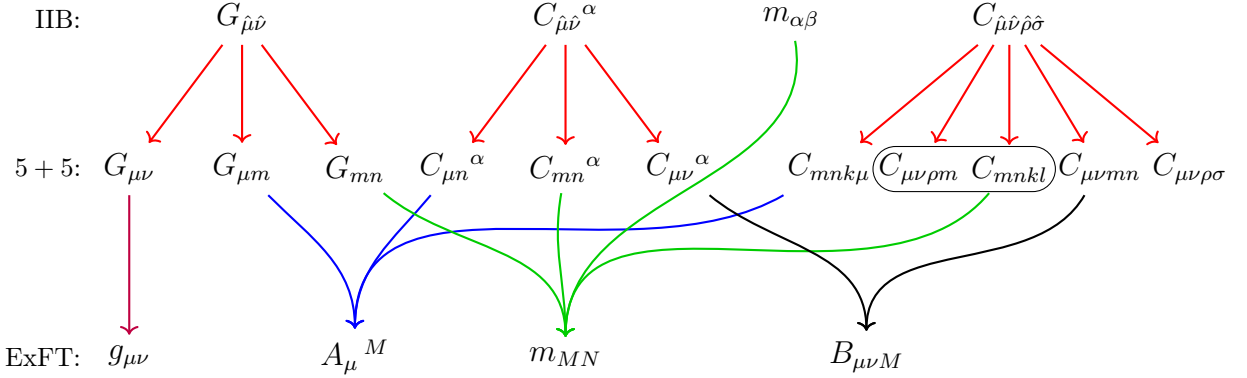
\begin{figure}[ht!]
	\centering
	\begin{tikzpicture}
	
    \node at (-12.3,7) (G) {$G_{\hat{\mu}\hat{\nu}} $};
	\node at (-8,7)    (C) {$C_{\hat{\mu}\hat{\nu}}{}^{\alpha}$};
    \node at (-5,7)   (mab) {$m_{\alpha \beta}$};
	\node at (-2.1,7) (Cf) {$C_{\hat{\mu}\hat{\nu}\hat{\rho}\hat{\sigma}}$};
   
	\node at (-13.8,5) (G1)  {$G_{\mu\nu}$};
	\node at (-12.3,5) (G2)  {$G_{\mu m}$};
	\node at (-10.8,5) (G3)  {$G_{mn}$};
	\node at (-8,5) (C1) {$C_{m n}{}^{\alpha}$};
	\node at (-9.5,5) (C2) {$C_{\mu n}{}^{\alpha}$};
	\node at (-6.5,5) (C3) {$C_{\mu \nu}{}^{\alpha}$};
    \node at (+0.3,5) (Cf0) {$C_{{\mu}{\nu}{\rho}{\sigma}}$};
    \node at (-2.1,5) (Cf4) {$C_{{m}{n}{k}{l}}$};
    \node at (-0.9,5) (Cf2) {$C_{\m\n{m}{n}}$};
    \node at (-4.5,5) (Cf3) {$C_{{m}{n}{k}{\mu}}$};
    \node at (-3.3,5) (Cf1) {$C_{\m\n\r{m}}$}; 

    \node at (-13.8,2.5) (GExFT) {$g_{\mu\nu}$};
	\node at (-10.8,2.5) (A) {$A_{\mu}\,^{M}$};
	\node at (-4,2.5) (B) {$B_{\mu \nu M}$};
	\node at (-8,2.5) (M) {$m_{MN}$};

    \draw[->, thick, red] (G) -- (G1);
	\draw[->, thick, red] (G) -- (G2);
	\draw[->, thick, red] (G) -- (G3);
 	\draw[->, thick, red] (C) -- (C1);
	\draw[->, thick, red] (C) -- (C2);
	\draw[->, thick, red] (C) -- (C3);
    
    \draw[->, thick, red] (Cf) -- (Cf0);
    \draw[->, thick, red] (Cf) -- (Cf4);
	\draw[->, thick, red] (Cf) -- (Cf2);
	\draw[->, thick, red] (Cf) -- (Cf3);
    \draw[->, thick, red] (Cf) -- (Cf1);

  	\draw[->, thick, purple] (G1) -- (GExFT);
	
	\draw[->, thick, blue] (G2) edge[out=-45,in=90] (A);
	\draw[->, thick, blue] (C2) edge[out=-130,in=90] (A);
    \draw[->, thick, blue] (Cf3) edge[out=-150,in=90] (A);

    \node at (-2.7,5.0) [rounded rectangle, draw, align=center, minimum width = 2.7cm, minimum height=0.6cm] (RNS) {};

	\draw[->, thick, green!80!black] (G3) edge[out=-40,in=90] (M);
    \draw[->, thick, green!80!black] (mab) edge[out=-80,in=90] (M);
	\draw[->, thick, green!80!black] (Cf4) edge[out=-130,in=90] (M);
    \draw[->, thick, green!80!black] (C1) edge[out=-100,in=90] (M);
	
	\draw[->, thick] (C3) edge[out=-40,in=90] (B);
    \draw[->, thick] (Cf2) edge[out=-120,in=90] (B);
	
    \node at (-14.3,7) [anchor=east] ()  {{ \footnotesize IIB:}};
    \node at (-14.3,5) [anchor=east] ()  {{ \footnotesize{$5+5$:}}};
    \node at (-14.3,2.5) [anchor=east] ()  {{ \footnotesize ExFT:}};

	\end{tikzpicture}
	\caption{Representation of the embedding of Type IIB into the $\rmE_{6(6)}$ ExFT. First layer of arrows (red) denotes the $10=5+5$ split of coordinates according to the standard KK ansatz. The bottom layer of arrows schematically shows how the result is combined into the fields of ExFT. The rounded rectangle combines fields that are dual. The 4-form $C_{\m\n\r\s}$ is non-dynamical and its field strength contributes to the embedding tensor of gauged supergravity.}
	\label{fig:split}
\end{figure}

In full details embedding of the Type IIB theory into the $\rmE_{6(6)}$ ExFT has been presented in \cite{Baguet:2015xha} (see also \cite{Hohm:2013vpa}) and is schematically represented in Fig \ref{fig:split}. The idea is that one starts with the SL(2)-covariant formulation of Type IIB supergravity and splits all ten-dimensional indices into 5+5 using the standard Kaluza-Klein ansatz, however, keeping full dependence on all the coordinates. Gauge fields must be dualized to the lowest possible rank of the form as in the prescription of \cite{Cremmer:1997ct}. The resulting fields can then be combined into irreps of the $\rmE_{6}$ symmetry group.

The expressions standing behind the bottom layer of arrows in Fig \ref{fig:split} are pretty complicated and include terms that mix the main contribution, that is where an arrow starts, with other fields. Since we are working in the truncated theory where $A_\m{}^M=0$, $B_{\m\n M}=0$ there is no need to dig into these details and the identification becomes pretty straightforward:
\begin{equation}
    \begin{aligned}
        G_{\mu \nu}&=|g|^{-\frac{1}{3}}g_{\mu \nu},  && G_{mn}= g_{mn}, \\
        B_{mn}{}^{\a}&=-\frac{1}{2}C_{mn}{}^{\a}, && C_{mnkl}=\ 4\, c_{mnkl},
    \end{aligned}
\end{equation}
where $g= \det g_{mn}$. To recover the deformed 4-form potential $C_{\m\n\r\s}$ we will also need the self-duality relation \eqref{FF55} that in the split notations takes the following form
\begin{equation}
\begin{aligned}
    &|g|^{\frac{1}{3}} \sqrt{-\det\left(g_{\mu \nu}(x_m,y_{\mu})\right)} \epsilon^{m n k l p}\left(\partial_{[m} C_{n k l p]}-\frac{3}{4}\epsilon_{\alpha \beta}C_{[mn}{}^{\alpha}\partial_{k} C_{l p]}{}^{\beta} \right)\\
    &=e^{-\phi}\sqrt{|g|} \sqrt{-\det\left(\bar{g}_{\mu \nu}(y_\m)\right)} \epsilon^{m n k l p}\left(\partial_{[m} C_{n k l p]}-\frac{3}{4}\epsilon_{\alpha \beta}C_{[mn}{}^{\alpha}\partial_{k} C_{l p]}{}^{\beta} \right)\\
    &=\epsilon^{\mu \nu \rho \sigma \tau}\partial_{[\mu} C_{\nu \rho \sigma \tau]}
\end{aligned}
\end{equation}
where in the second line we performed the rescaling \eqref{renorm_formulas}. In terms of the 10-dimensional fields of the Type IIB theory the introduced truncation can be summarized as follows
\begin{equation}
    \label{eq:bg10truncated}
    \begin{aligned}
        ds_{10}^2 & = g^{-\fr13}g_{\m\n}(y,x)dy^\m dy^\n + g_{mn}(x)dx^m dx^n ,\\
        C_4 & = \fr1{4!}C_{\m_1\dots \m_4}dy^{\m_1}\wedge \dots \wedge dy^{\m_4} + \fr1{4!}C_{m_1\dots m_4}dx^{m_1}\wedge \dots \wedge dx^{m_4}.
    \end{aligned}
\end{equation}
And on top of that the 5-form field strength must be self-dual in ten dimensions. Deformed backgrounds will also follow the same ansatz.

\section{Deformations in the truncated Type IIB theory}
\label{sec:deform}

Let us now proceed with polyvector deformations of Type IIB backgrounds using the formalism of the $\rmE_{6(6)}$ exceptional field theory. Following the prescription introduced in \cite{Gubarev:2020ydf} we construct polyvector deformations as exponents of generators of the $\rmE_{6(6)}$ algebra of negative level with respect to the geometric $\mathfrak{gl}(1)$. From \eqref{split27B} we find that these are
\begin{equation}
    T_{mn}{}^\a, \quad T_{mnkl},
\end{equation}
and hence deformations will be parameterised by a doubled of bi-vectors $\b_\a{}^{mn}$ and by a 4-vector $\W^{mnkl}$. Certainly, one component of the doublet is the familiar bi-vector  Yang-Baxter deformation that has been intensively studied in the literature in various contexts (see e.g. reviews \cite{vanTongeren:2013gva,Orlando:2019his,Seibold:2020ouf} and references therein). The general form of a deformation can be written in terms of the rescaled generalized vielbein as follows
\begin{equation}
    E'_M{}^A = O_M{}^N E_N{}^A.
\end{equation}
The matrix $O \in \rmE_{6(6)}$ is generated by $T_{mn}{}^\a$, $T_{mnkl}$ contracted with polyvectors and is most conveniently represented as a product
\begin{equation}
    \label{eq:defmatrix}
    \left(O\right)_{M}{}^{N}=\left(O^2 \, O^4\right)_{M}{}^{N},
\end{equation}
where the matrices $O^2$ and $O^4$ are given by
\begin{align}
 \left(O^2\right)_{M}{}^{N}&=\exp\left(T^\gamma{}_{mn} \beta_{\gamma}{}^{mn}\right)_{M}{}^{N}\nonumber
 \\ 
 &= \begin{pmatrix}
        \delta_{m}{}^{n} &  0 & 0 & 0 
        \\
         -2\beta_\gamma{}^{ m n}\epsilon^{\gamma \alpha} &  \delta_{n}{}^{m} \delta_{\beta}{}^{\alpha} & 0 & 0 
        \\
         -\sqrt{2}\beta_\gamma{}^{ m n}\widetilde{\beta}^\gamma{}_{ m m_1 m_2}  & \sqrt{2}\,\epsilon_{ \alpha\beta}\widetilde{\beta}^\alpha{}_{ n m_1 m_2} & \delta_{m_1 m_2}^{n_1 n_2} & 0 
        \\
         \frac{2}{3}\beta_\gamma{}^{ m n}\widetilde{\beta}^\gamma{}_{ m n_1 n_2}\beta_\beta{}^{ n_1 n_2} \epsilon^{\beta \alpha}&\beta_\gamma{}^{ n_1 n_2} \epsilon^{\gamma \alpha} \,\epsilon_{\beta \delta}\widetilde{\beta}^{\delta}{}_{ n n_1 n_2}& \sqrt{2}\, \epsilon^{\alpha\beta } \beta_\beta{}^{ n_1 n_2}  &  \delta_{\beta}{}^{\alpha} 
    \end{pmatrix},
 \label{tens_def_2_rang} \\
 \left(O^4\right)_{M}{}^{N}&= \exp\left(T_{m_1 \ldots m_4} \Omega^{m_1 \ldots m_4}\right)_{M}{}^{N}= \begin{pmatrix}
        \delta_{m}{}^{n} &  0 & 0 &  0 \\
         0 &  \delta_{n}{}^{m} \delta_{\beta}{}^{\alpha} & 0 &  0\\
         \sqrt{2}\, \tW_{[m_1} \d_{m_2]}{}^n &  0 & \delta_{m_1 m_2}^{n_1 n_2} &  0 \\
         0 & -\delta_\beta{}^\alpha\, \tW_ n & 0 & \delta_{\beta}{}^{\alpha}
    \end{pmatrix}
 \label{tens_def_4_rang}.
 \end{align}
To make expressions more clear we introduce the following notations
 \begin{equation}
 \begin{aligned}
 & \widetilde{\beta}^\alpha{}_{ m n_1 n_2}=\frac{1}{2}\,\epsilon^{\alpha\beta}\,\epsilon_{ m  n_1 n_2  p_1 p_2}\,\beta_\beta{}^{ p_1 p_2}\,, &&
 \tW_ m  = \frac{1}{4!}\,\epsilon_{ m  p_1\cdots  p_4}\,\Omega^{ p_1\cdots  p_4}\,,
  \end{aligned}
\end{equation}

\subsection{4-vector transformation rules}

In the case of vanishing 4-vector parameter  one ends up with a deformation generated by an $\rmSL(2)$ doubled of bi-vectors. These can be obtained from the standard bi-vector Yang--Baxter deformation by an action of the $\rmSL(2)$ S-duality group. Therefore, to arrive at a generalization of bi-vector deformations in the Type IIB case one gets most in deformations that include 4-vectors. Also when speaking of deformations along compact isometries from general grounds and comparing to the generalized Type IIB Yang-Baxter equations of \cite{Sakatani:2020wah} one concludes that in the absence of 4-vectors, the doublet of bi-vectors is restricted by a doublet of classical Yang-Baxter equations, when the bi-Killing ansatz is imposed:
\begin{equation}
    \b_\a{}^{mn} = r_\a{}^{i_1 i_2}k_{i_1}{}^mk_{i_2}{}^n.
\end{equation}
However, the theorem of \cite{Lichnerowicz:1988abc} forbids non-abelian deformations and hence one is left only with commuting Killing vectors $k_i{}^m$ on the sphere. Therefore, the first component of the doublet generates the standard TsT transformation of \cite{Lunin:2005jy}, while the second gives its S-dual.

To reduce technicality and to end up with a nice example, instead of going to the most general case when both the bi-vectors and the 4-vector are non-zero, we consider only 4-vector deformations. In this case the transformation of the generalized metric becomes
\begin{equation}
    {\mM'}_{ M N}= \left(O^4\right)_{M}^{\ \ M'}\left(O^4\right)_{N}^{\ \ N'}{\mM}_{ M' N'}
\end{equation}
and given the truncation the rescaled generalized metric is takes following form
\begin{equation}
\label{M2B_re}
{M}_{MN} =e^{2\phi} \left(
\begin{array}{cccc}
{M}_{mn}& 0 &{M}_{m,n_1 n_2}& 0\\
0 &{M}^{m\alpha,}{}^{n\beta}& 0&{M}^{m\alpha,}{}^{\beta}\\
{M}_{m_1 m_2,n}&0&{M}_{m_1 m_2,n_1 n_2}&0\\
0&{M}^{\alpha,n\beta}&0&{M}^\alpha{}^{\beta}
\end{array}
\right)
\;.
\end{equation}
where
\begin{equation}
\begin{aligned}
    M_{m n}&=\left(\pm g_{m n} + \frac{1}{2}  \tc^{n_1 n_2}_{m}  g_{n_1 n_2, \, m_1' m_2'}\tc^{m_1' m_2'}_{n}\right) 
\\
    M_{m_1 m_2, n}&=  \frac{1}{\sqrt{2}}\left| g \right|^{-\frac{1}{2}}g_{m_1 m_2, \, m_1' m_2'}\tc^{m1' m2'}_{n} 
\\
    M^{\alpha m,\beta n}&=   m^{\alpha \beta} \left(g^{m n} \pm  \tc^{m } \tc^{n }\right)
\\
    M^{\alpha,n \beta}&=   \mp \left| g \right|^{-\frac{1}{2}}m^{\alpha \beta}\tc^{n }
\\
    M_{m_1 m_2, n_1 n_2}&=\left| g \right|^{-1}g_{m_1 m_2, \, n_1 n_2}  
\\
    M^{\alpha \beta}&=
 \pm \left| g \right|^{-1}m^{\alpha \beta},
\end{aligned}
\label{our_generalized_metric}
\end{equation}
and again the upper/lower signs correspond to cases with Euclidean/Minkowskian signature of the metric $g_{mn}$. 

By a direct comparison of blocks of the initial generalized metric \eqref{M2B_re} and the deformed one we derive the following deformation rules for (components of) the Type IIB supergravity fields
\begin{equation}
\begin{aligned}
     K&= \left(1+W \tc\right)^2 \pm W^2\\
    {m'}^{\alpha \beta}&=m^{\alpha \beta}\\
   G'_{\mu \nu}&=K^{\frac{1}{2}} G_{\mu \nu}\\
    \tc'{}^{m }&=K^{-\frac{1}{4}} \tc^{m }\left(1+ W  \tc \right) \pm K^{-\frac{1}{4}}W^m  \\
    g'_{m n} &= K^{-\fr12} \left(g_{m n}
        +2\, W_{(m}\tc_{n)} \pm \left(1\pm\tc^2\right) W_m W_n \right),\\
    W_m&=\frac{1}{4!}\varepsilon_{m p_1 \ldots p_4} \tW^{ p_1 \ldots p_4}, \\
    g' & = K^{-\fr32}g,
\label{deformed_IIB_fields}
\end{aligned}
\end{equation}
where indices on the RHS are raised and lowered by the undeformed metric $g_{mn}$, and the upper(lower) sign corresponds to Euclidean(Minkowskian) signature of the metric $g_{mn}$. These expressions are very similar to that of \cite{Bakhmatov:2019dow} representing tri-vector deformations of 11D backgrounds in the 11=7+4 split withing the SL(5) exceptional field theory. The reason is that in both these cases rank of the deformation tensor and of the non-vanishing gauge field is one less than dimension of the ``internal'' space, and hence by contraction with epsilon tensor it can be converted into a 1-form. Note that the SL(2) metric $m_{\a\b}$ and the external metric $\bar{g}_{\m\n}$ that are scalars under $\rmE_{6(6)}$ do not transform under the deformation.

It is also convenient to list explicitly transformation rules for the internal and external components of the self-dual 5-form field strength. The former is obtained by simply taking a covariant derivative of $\tc^m$ and reads
\begin{equation}
    \label{eq:deform_int}
    F_{int}' = \fr{1}{5!} \dt_{m}\left(K^{-\fr{3}{4}}g^{\fr12}\tc'{}^m\right) \e_{n_1\dots n_5} dx^{n_1}\wedge\dots \wedge dx^{n_5}.
\end{equation}
Note that $\e_{n_1\dots n_5}$ is the epsilon symbol. One may substitute the $\tc'^{m}$ from \eqref{deformed_IIB_fields} in the expression above and write the transformation rule expression new field strength in terms of the old one, however in practice it is more convenient to first derive $\tc'{}^m$ and then calculate $F'_{int}$. The external components of the flux are then restored from the self-duality condition and read
\begin{equation}
    \label{eq:deform_ext}
    F_{ext}' = K^2g^{-\fr12}\dt_m\left(K^{-\fr{3}{4}}g^{\fr12}\tc'{}^m\right) \fr{1}{5!}\sqrt{-\det G_{\m\n}}\e_{\m_1 \dots \m_5}dx^{\m_1}\wedge \dots \wedge dx^{\m_5}.
\end{equation}
Note using both $g$ and $g'$ in the expression

\subsection{Transformation of fluxes}

Certainly, an arbitrary transformation of the fields of the form \eqref{deformed_IIB_fields} will not map a solution to Type IIB equations again into a solution. As it has been advocated in \cite{Gubarev:2020ydf} the simplest way to derive conditions on $\tW_m$ for the deformation to be a solution generating transformation is to require invariance of generalized fluxes $\mF_{AB}{}^C$. Note, that this also makes perfect sense from the algebraic point of view simply stating that $\mF_{AB}{}^C$ are scalars under $\rmE_{6(6)}$ as they should be since the group acts only on curved indices. Recall components of the generalised flux in the $\rmE_{6(6)}$ exceptional field theory
\begin{equation}\label{Fstructure1}
    \mF_{A,B}{}^{C} = 2 E_{M}{}^{C} E^{K}{}_{[A|} \dt_K E^{M}{}_{|B]} + 10 d^{MKR} d_{NLR} E_{M}{}^{C} E^{N}{}_{B} \dt_K E^{L}{}_{A}.
\end{equation}
It is worth noting, that in contrast to the generalised Scherk--Shwarz reduction of \cite{Musaev:2013rq} these are not required to be constants. Quite the opposite, generalized fluxes are simply a convenient way to pack generalized vielbeins. Under the $\rmE_{6(6)}$ group the generalized flux decomposes into irreducible representations $\mF_{A,B}{}^{C} \in \bf{27} \, \oplus \, \bf{351}$. In components one has the trombone $\theta_{A} \in \bf{27}$ and  Z-flux $Z_{AB}{}^C \in \bf{351}$ \cite{Musaev:2013rq}
\begin{equation}
    \begin{aligned}
       -27 \theta_{A} & = 9 \partial_{M} E^{M}{}_{A} + E_{M}{}^{B} \partial_{A} E^{M}{}_{B}, \\
       Z_{A B}{}^{C} & = {}5\left({e^{-3 \f}e} d_{A B E} d^{N M K} E_{N}\,^{C} \partial_{M}{E_{K}\,^{E}}+d_{A B E} d^{N M K} E_{N}\,^{C} E_{K}\,^{E} \partial_{M}\left(e^{-3 \f}e\right)\right),
\end{aligned}
\label{all_flux_fields}
\end{equation}
where we denote  $\partial_{A}=E_{A}{}^{K}\partial_{K}$. The additional factors of $e^{-3 \f}$ come from turning from curved to flat indices in the invariant tensor
\begin{equation}
\label{DtensorsEtrunc}
    d_{M N P} E^{M}{}_{A} E^{N}{}_{B} E^{P}{}_{C} = e^{-3 \phi} \, e \,d_{A B C}.
\end{equation}
For components of the latter one has
\begin{equation}\begin{aligned}
 & d_{ABC} : &&
 d_{a}{}^{b \overline{\alpha},\overline{\beta}} = \frac{1}{\sqrt{10}}\, \delta^b_a \epsilon^{\overline{\alpha}\overline{\beta}}\;,&& 
d_{a b}{}^{c \overline{\alpha},d \overline{\beta}} = \frac1{\sqrt{5}}\, \delta_{ab}^{cd}\,\epsilon^{\overline{\alpha}\overline{\beta}}\;, &&
d_{ab,cd,e} = \frac1{\sqrt{40}}\,\epsilon_{abcde}\;.
 \label{dtens}
\end{aligned}
\end{equation}

To derive an algebraic condition that can be interpreted as a generalisation of the classical Yang--Baxter equation we restrict the deformation tensor by the poly-Killing ansatz
\begin{equation}
    \Omega^{m n k l}=\rho^{i_1 i_2 i_3 i_4}k^{m}_{i_1}k^{n}_{i_2}k^{k}_{i_3}k^{l}_{i_4}
\end{equation}
where $i_1,i_2,...$ label Killing vectors $k^{m}_{i}$ of the initial Type IIB solution. These form an algebra of isometries under the standard Lie bracket
\begin{equation}
    [k_{i_1},k_{i_2}] = f_{i_1i_2}{}^{i_3}k_{i_3}.
\end{equation}
The generalized  classical $r$-matrix $\r^{i_1i_2 i_3i_4}$ is conventionally taken to be constant and completely antisymmetric. Since we require covariance of the generalized flux under a 4-vector transformation it is convenient to single out a non-covariant term and require it to vanish. Since $\mF_{AB}{}^C$ must be a scalar the non-covariant piece is defined simply as
\begin{equation}
    \d \mF_{AB}{}^C = \mF'_{AB}{}^C - \mF_{AB}{}^C.
\end{equation}
This gives the following sufficient conditions on $\r^{i_1 i_2 i_3 i_4}$ 
\begin{align}
    3\rho^{i_1 i_4 [i_2 i_3 } \rho^{i_5 i_6 |i_7 i_8|}   
    f_{i_1 i_7}{}^{i_9]}+ \rho^{i_1i_4 [i_2i_3 } \rho^{i_5 i_6  i_9] i_7}   f_{i_1 i_7}{}^{i_8}=0 \label{four_gcybe}\\
    \rho^{ i_1  i_4 i_2 [i_3} f_{i_1 i_4}{}^{i_5]}=0.
    \label{four_unimod}
\end{align}
This is a straightforward but lengthy calculation done mainly by the Cadabra 2 computer algebra program \cite{Peeters:2007wn}. Some details of  the calculation are presented in Appendix \ref{app:genfluxtransform}, and the Cadabra 2 files containing full calculations can be found in \cite{cadabra2IIBcode}. The first equation \eqref{four_gcybe} is a generalization of the classical Yang--Baxter equation for the 4-vector case. It has the form similar to that of the equation for a 3-vector deformation of \cite{Sakatani:2019zrs}. The second equation \eqref{four_unimod} is a generalization of the unimodularity condition. Abandoning it while keeping the first equation satisfied will generate solutions to a set of generalized 10D Type IIB equations along the lines of \cite{Bakhmatov:2022rjn,Bakhmatov:2022lin}. It is worth to mention that the relation between generalized flux invariance and the algebraic equations \eqref{four_gcybe}, \eqref{four_unimod} is heavily based on the condition that the geometric vielbein $e_m{}^a$ vanished under Lie derivative along Killing vectors of the metric and the gauge fields.

\subsection{Type IIB Exceptional Drinfeld  algebra}

The familiar classical Yang--Baxter equation can be understood as the condition for a certain transformation of the classical Drinfeld double to preserve its structure. Generalized Yang--Baxter equations for generalized $r$-matrices of any rank can be understood similarly from the point of view of exceptional Drinfeld algebras (EDA). The latter were first introduced in \cite{Sakatani:2019zrs,Malek:2019xrf} together with a generalization of the classical Yang-Baxter equation. The same equation as a sufficient condition for a 3-vector deformation to generate solutions to 11D supergravity equations has been derived earlier in \cite{Bakhmatov:2019dow} in the form of vanishing R-flux. Note however, that a sufficient condition that is extracted from the non-covariant part of the generalized flux tCransformation might be of various forms that differ by terms identically vanishing due to antisymmetrizations and various contractions. Hence, it is necessary to check that the equations \eqref{four_gcybe} indeed have the proper algebraic origin.

For that we start with the $\rmE_{6(6)}$ EDA in the Type IIB parametrization constructed in \cite{Lee:2016qwn}. This is a 27-dimensional Leibniz algebra whose generators $T_A$ can be labelled as 
\begin{align}
 (T_A) =\bigl(T_{a},\, T_\alpha{}^{a},\, \frac{1}{\sqrt{3!}}T^{a_{1}a_{2} a_{3}},\, \frac{1}{\sqrt{5!}}T_\alpha{}^{a_{1}\cdots a_{5}}\, \bigr)\,,
\label{Sac_gen}
\end{align}
where $a \in (1...5), \alpha \in (1...2)$. In the context of M-theory this algebra is a symmetry of equations of motion of the membrane on a 5-dimensional group manifold generated e.g. by $\{T_a\}$.  The multiplication table for this algebra is given by
\begin{equation}
\begin{aligned}
 T_{a}\circ T_{ b} &=f_{a b}{}^{c}\,T_{c}\,,
\\
 T_{a}\circ T_\beta{}^{ b} &= f_{\beta, a}{}^{c b}\,T_{c} 
 + f_{a,\beta}{}^{\gamma}\,T_\gamma{}^{ b} - f_{a c}{}^{ b}\,T_\beta{}^{ c},
 \\
 T_\alpha{}^{a}\circ T_{ b} &= 
 f_{\a, b}{}^{a c} \, T_{ c} 
 + 2\,\delta_{[ b}{}^{a}f_{ c],\alpha}{}^{\gamma}\, T_\gamma{}^{ c}
 + f_{ b c}{}^{a}\,T_\alpha{}^{ c} ,
\\
 T_{a}\circ T^{ b_1 b_2 b_3} &=
 f_{a}{}^{ c b_1 b_2 b_3}\, T_{ c} + 3\,\epsilon^{\gamma\delta}\,f_{\g, a}{}^{[ b_1 b_2}\, T_{\delta}^{ b_3]} - 3\,f_{a c}{}^{[ b_1}\, T^{ b_2 b_3] c} ,
 \\
 T^{a_1a_2a_3} \circ T_{ b} 
 &= -f_{ b}{}^{ c a_1a_2a_3}\,T_{ c} 
 - 6\,\epsilon^{\gamma\delta}\,f_{\g,[ b|}{}^{[a_1a_2}\delta_{| c]}{}^{a_3]}T_\delta^{ c}+ 3\,f_{ b c}{}^{[a_1}\, T^{a_2a_3] c}
\\
 T_{a}\circ T_\beta{}^{ b_1\cdots  b_5} &=
 5\, f_{a}{}^{[ b_1\cdots  b_4}\,T_\beta{}^{ b_5]}
 - 10\, f_{\b,a}{}^{[ b_1 b_2}\,T^{ b_3 b_4 b_5]} 
 + f_{a,\beta}{}^{\gamma}\, T_\gamma{}^{ b_1\cdots  b_5} 
 - 5\,f_{a c}{}^{[ b_1}\,T_\beta^{ b_2\cdots  b_5] c} ,
\\
 T_\alpha{}^{a_1\cdots a_5} \circ T_{ b} 
 &= - 10\,f_{[ b}{}^{[a_1\cdots a_4}\, \delta_{ c]}{}^{a_5]}T_\alpha{}^{ c}
 - 30\,f_{\a, c}{}^{[a_1a_2}\, \delta_{ b}{}^{a_3}\, T^{a_4a_5] c} 
 + 10\,f_{\a, b}{}^{[a_1a_2}\,T^{a_3a_4a_5]}\\
 &\quad \, + 5\,f_{ c, \alpha}{}^{\gamma}\, \delta_{ b}{}^{[a_1} T_\gamma{}^{a_2\cdots a_5] c}
 - f_{ b, \alpha}{}^{\gamma}\,T_\gamma{}^{a_1\cdots a_5}
  + 5\,f_{ b c}{}^{[a_1}\, T_\alpha{}^{a_2\cdots a_5] c}
\\
 T_\alpha{}^{a}\circ T_\beta{}^{ b} &= - f_{\a, c}{}^{a b}T_\beta{}^{ c} 
 - f_{ c, \alpha}{}^\gamma\,\epsilon_{\gamma\beta}\,T^{ c a b} 
 + \frac{1}{2}\,\epsilon_{\alpha\beta}\,f_{ c_1 c_2}{}^{a}\,T^{ c_1 c_2 b},
\\
 T_\alpha{}^{a}\circ T^{ b_1 b_2 b_3}
 &= -3\,f_{\a, c}{}^{a[ b_1}T^{ b_2 b_3] c} 
  - f_{ c, \alpha}{}^\gamma\,T_\gamma{}^{a c  b_1 b_2 b_3} 
  - \frac{1}{2}\,f_{ c_1 c_2}{}^{a}\,T_\alpha{}^{ c_1 c_2  b_1 b_2 b_3},
\\
 T^{a_1a_2a_3} \circ T_\beta{}^{ b} 
 &= -f_{ c}{}^{a_1a_2a_3  b}T_\beta{}^{ c}
  + 3\, f_{\b, c}{}^{[a_1a_2}\,T^{a_3]  b c}
  + \frac{3}{2}\, f_{ c_1 c_2}{}^{[a_1}\,T_\beta{}^{a_2a_3]  b  c_1 c_2},
\\
 T_\alpha{}^{a}\circ T_\beta{}^{ b_1\cdots  b_5}
 &= -5\, f_{\a, c}{}^{a[ b_1}\,T_\beta{}^{ b_2\cdots  b_5] c} 
\\
 &\quad \, + 3\, f_{ c_1 c_2}{}^{[a_1}\,\delta_{ b}{}^{a_2}T^{a_3] c_1 c_2},
\\
 T_\alpha{}^{a_1\cdots a_5} \circ T_\beta{}^{ b} 
 &= -5\,\epsilon_{\alpha\beta}\,f_{ c}{}^{[a_1\cdots a_4}\,T^{a_5]  b c}
 + 10\,f_{\a, c}{}^{[a_1a_2}\,T_\beta{}^{a_3a_4a_5]  b c} \,,
\\
 T^{a_1a_2a_3} \circ T^{ b_1 b_2 b_3} 
 &= -3\, f_{ c}{}^{a_1a_2a_3 [ b_1}\, T^{ b_2 b_3]  c}
   + 3\, \epsilon^{\gamma\delta}\,f_{\g, c}{}^{[a_1a_2}\,T_\delta{}^{a_3] b_1 b_2 b_3  c} \,,
\\
 T^{a_1a_2a_3} \circ T_\beta{}^{ b_1\cdots  b_5} 
 &= -5\,f_{ c}{}^{a_1a_2a_3 [ b_1}\,T_\beta{}^{ b_2\cdots  b_5]  c} \,,
\\
 &\quad\,
 + 10\,f_{ c_1 c_2}{}^{[a_1}\, \delta_{ b}{}^{a_2}\,T_\alpha{}^{a_3a_4a_5] c_1 c_2},
\\
 T_\alpha{}^{a_1\cdots a_5} \circ T^{ b_1 b_2 b_3} 
 &= 5\,f_{ c}{}^{[a_1\cdots a_4}\, T_\alpha{}^{a_5]  b_1 b_2 b_3 c} ,
\end{aligned}
\label{eda_lgebra}
\end{equation}
and all other products vanish. In a compact form this table can be written as
\begin{equation}
    T_A\circ T_B = \bar{\mF}_{AB}{}^C T_C,
\end{equation}
where $\bar{\mF}_{AB}{}^C $ carries the same meaning as the generalized flux $\mF_{AB}{}^C$ but now it is constant at has only non-vanishing components determined by
\begin{equation}
    \begin{aligned}
        & f_{ab}{}^c, && f_{\a a}{}^{bc}, && f_{a}{}^{bcde}, && f_{a,\a}{}^\b.
    \end{aligned}
\end{equation}

The idea of the deformation of an EDA is that one starts, say, with a simple EDA with the only non-vanishing structure constant being $f_{ab}{}^c$. Then one performs a transformation of generators according to
\begin{equation}
    T_A \to O_A{}^B T_B,
\end{equation}
where $O_A{}^B$ is precisely the deformation matrix \eqref{eq:defmatrix} with Killing vectors being simply Kronecker delta tensors. This is due to the fact that we are on a group manifold on which the group acts transitively and adapted coordinates might be chosen. The requirement is then that the multiplication table generated by such a transformation is in accordance with \eqref{eda_lgebra}. Keeping only the 4-vector deformation we end up with the following condition (see \cite{Lee:2016qwn})
\begin{align}
    -2f_{{c} {d}}\,^{{[b_1}}\rho^{ {b_2} {b_3} {b_4}] c } \rho^{ d a_1 a_2 a_3} 
    +3 f_{{c} {d}}\,^{{[a_1}}\rho^{ {a_2} {a_3}] {d} [{b_1} } \rho^{ {b_2}  {b_3} {b_4}] {c}}   =0 \label{four_gcybe_sak}\\
    \rho^{ {a_1}  {a_2} {b} {c}} f_{{a_1} {a_2}}\,^{{d}}=0
    \label{four_unimod_sak}
\end{align}
where $a, b, c, d = 1,...,5$. 

Although at the first glance these conditions look different from those obtained from exceptional field theory and listed in \eqref{four_gcybe}, they are actually equivalent on a group manifold. Indeed, let us define 
\begin{equation}
    \rho^{a b c d}=\varepsilon^{a b c d e} \rho_{e},
\label{4_to_1}
\end{equation}
that is possible only in the case when the total amount of Killing vectors is equal to dimension of the background space (that is five). In terms of $\r_a$ both \eqref{four_gcybe_sak}, \eqref{four_unimod_sak} and \eqref{four_gcybe}, \eqref{four_unimod} take the following particularly simple form 
\begin{align}
    \r_d\rho_{[a} f_{{b]} {c}}\,^{{c}}=0 \label{one_gcybe}\\
    \rho_a f_{{b} {c}}\,^{{a}}=0.
    \label{one_unimod}
\end{align}
Note, that this is a non-trivial check, as the LHS of above equations are not the only expressions one is able to compose from $\r_a$ and $f_{ab}{}^c$. Therefore, we conclude that the generalized classical Yang-Baxter equation \eqref{four_gcybe} are indeed the conditions for a deformation to preserve the structure of an $\rmE_{6(6)}$ EDA in the Type IIB parametrisation.

\section{Example of 4-vector deformation of \texorpdfstring{$\AdS_5\times  \SS^5$}{AdS5xS5}}
\label{sec:example}

As an explicit example of deformations let us consider a 4-vector generalization of the DPP deformation of \cite{Bakhmatov:2020kul} by multiplying the 3-vector DPP used  there by an additional momentum generator. Therefore, as in the 3-vector case this deformation satisfies the generalized Yang--Baxter equation \eqref{four_gcybe} and is non-unimodular, i.e. does not satisfy \eqref{four_unimod}. However, under certain additional condition on the deformation parameters the deformed background is a solution to the standard Type IIB supergravity equation. 

Using Killing vectors of $\AdS_5$ in the Poincar\'e patch 
\begin{equation}
\begin{aligned}
&&P_{a} &= \dt_a, & K_a &= x^2 \dt_{a} + 2 x_a D,\\
&&D&=-x^m\dt_m, &  M_{ab} &= x_a \dt_b - x_b \dt_a,
\end{aligned}
\end{equation}
we obtain the following DPPP deformation tensor 
\begin{equation}
\label{dpp-omega}
\begin{aligned}
    \W & = \frac{\r_{a}}{2R}  \e^{abcd}\, D\wedge P_{b} \wedge P_{c} \wedge P_{d} \\
    &= -\frac{\r_{a}}{2 R}\, \left( x^{a}\, \dt_0 \wedge \dt_1 \wedge \dt_2 \wedge \dt_3 + \, z\,  \e^{abcd}\, \dt_{b} \wedge \dt_{c} \wedge \dt_{d} \wedge \dt_z\right).
\end{aligned}
\end{equation}
Here $a,b=0,1,2,3$ and $m,n=0,1,2,3,z$, and we used definition $x^2 = \h_{mn} x^m x^n$ and $x_a = \h_{ab}x^b$.
For the 1-form $W_mdx^m$ and the scalar function $K$ we obtain
\begin{equation}
    \begin{aligned}
        W &=\fr12 \left(\frac{R}{z}\right)^4 \rho_{a} \left( dx^a - x^a \frac{dz}{z}\right),\\
        K & =1-\left(\frac{R}{z}\right)^3  \frac{\r_ax^a}{z}.
     \end{aligned}
\end{equation}
Using the transformation rules \eqref{deformed_IIB_fields} we obtain the following deformed background
\begin{equation}
\label{eq:ads5s4deformed}
\begin{aligned}
ds^2 &= K^{-\fr12} \left(\frac{R}{z}\right)^2  \left[-(dx^0)^2+(dx^1)^2+(dx^2)^2 +K \,dz^2 +\fr12 \left(\frac{R}{z}\right)^3\r_a dx^a dz \right]\\
&+ {R^2} K^{\fr12} d\W_{(5)}^2,\\
F &=- \fr1R \left( -K^{-2} d\text{Vol}(AdS_5)+d\text{Vol}(\SS^5)\right).
\end{aligned}
\end{equation}
For this to be a solution to Type IIB supergravity equations $\r^{a}$ must be restricted by $\r^2=-(\r^0)^2+(\r^1)^2+(\r^2)^2+(\r^3)^2=0$. As for the DPP deformation one could expect that this background would be a solution to a generalization of Type IIB supergravity equation similar to those found in \cite{Bakhmatov:2020kul}, but with an additional tensor $I^{[mn],k}$ instead of the Killing vector $I^m$.

\section{Conclusions}
\label{sec:conclusions}

In this work we consider polyvector deformations of Type IIB backgrounds by embedding of the Type IIB supergravity theory into the $\rmE_{6(6)}$ exceptional field theory and performing $E_{6(6)}$ transformations parameterized by a 4-vector and an $\rmSL(2)$ doublet of bi-vectors. Taken in a poly-Killing ansatz these are parameterized by generalized $r$-matrices $\r^{i_1i_2i_3i_4} $ and $\r^{\a,i_1i_2}$, where $\a=1,2$ labels components of the doublet and $i_m$ label Killing vectors. Due to the embedding these Killing vectors can only be taken along isometries of a five-dimensional submanifold of a full 10-dimensional solution referred to as the internal, although no KK reduction is assumed. For simplicity and having in mind the example of $\AdS_5\times \SS^5$ we restrict the formalism to only backgrounds of the form $M_5\times N_5$, meaning that there are no off-block-diagonal components in the full 10D metric, metric on $N_5$ does not depend on coordinates on $M_5$ and vice versa. The self-dual 5-form field strength has components both along $M_5$ and along $N_5$, which are however proportional to the corresponding volume form and hence fit the truncation. This significantly simplifies the formalism of the corresponding exceptional field theory reducing it effectively to dynamics of the (rescaled) generalized metric $M_{MN}$ and the external metric $g_{\m\n}$. Explicitly the relation between Type IIB fields and the generalized metric is given in \eqref{M2B_re}, \eqref{our_generalized_metric}. Backgrounds fitting the truncation are of the form \eqref{eq:bg10truncated}. 

The most general polyvector deformation of a Type IIB background is given by the matrices \eqref{tens_def_2_rang}, \eqref{tens_def_4_rang}. Taken in the poly-Killing ansatz these generate solutions only if a certain generalization of the classical Yang-Baxter equation and of the unimodularity constraint hold. For 4-vector deformations we find them explicitly and list in \eqref{four_gcybe} and \eqref{four_unimod}. As expected these are precisely the same conditions for a 4-vector deformation of the corresponding Type IIB exceptional Drinfeld algebra to preserve its structure presented in \cite{Lee:2016qwn}. In the absence of the 4-vector the latter are simply a doublet of the ordinary classical Yang--Baxter equations and hence can be obtained by an S-duality rotation of a usual bi-vector Yang--Baxter deformed solution. Moreover, due to \cite{Lichnerowicz:1988abc,Pop:2007abc} these have no non-abelian solutions on compact isometries.  Deformations along isometries of the compact space of solutions of the type $\AdS \times N$ correspond to exactly marginal deformations of the dual theory and abelian deformations are simply TsT. For this reason we do not consider bi-vectors without 4-vectors, while keeping them both ends up with too complicated Yang--Baxter equations. Instead, we focus only at 4-vector deformations and provide explicit transformation rules for supergravity fields \eqref{deformed_IIB_fields}. We construct an example for the $\AdS_5\times \SS^5$ background that is based on the 3-vector solution found in \cite{Bakhmatov:2020kul}. Adding a $\rmU(1)$ generator we end up with a 4-vector and derive a deformed solution \eqref{eq:ads5s4deformed}. Similarly to the example of \cite{Bakhmatov:2020kul} this background should not preserve any of the original supersymmetries, however we didn't check it explicitly.

There are several directions of extending the presented results that we find interesting. Certainly, one is interested in finding a 4-vector deformation that does not split as $\W_3 \times k_\bullet$ into a 3-vector and a $\rmU(1)$ isometry, as well as solutions with both the doublet of bi-vectors and the 4-vector non-vanishing. It would be interesting to construct 4-vector deformations along isometries of the sphere $\SS^5$ and to investigate holographic duals of these solutions, that are presumably new exactly marginal deformations of $\mc{N}=4$ $D=4$ SYM and of the presented here DPPP deformation. The latter must be a non-commutative deformation of the type analyzed in \cite{vanTongeren:2015uha}.

To keep matters simple we introduced a truncation of the full $\rmE_{6(6)}$ ExFT, that significantly restricts the set of backgrounds allowed. Therefore it would be useful to develop the most general approach with all deformations and all fields non-vanishing. This would require to abandon the truncation, to construct a full flux formulation of the $\rmE_{6(6)}$ theory along the lines of \cite{Gubarev:2023kvq} and to check invariance of the fluxes.

Dropping the unimodularity condition \eqref{four_unimod} while keeping the generalized Yang--Baxter equation \eqref{four_gcybe} leads to backgrounds that satisfy a more general set of equations \cite{Bakhmatov:2022rjn}. In the case of bi-vector Yang--Baxter non-unimodular deformations one ends up with the generalized 10D supergravity equation constructed in \cite{Arutyunov:2015mqj}. This theory has an additional vector $I^m$ required to be a Killing vector of any solution to field equations. The Green--Schwarz superstring is known to be kappa-symmetric on solutions to this generalized set of equations with the vector $I^m$ appearing in the dimension $\fr12$ component of torsion constraints. Therefore, it appears to be interesting to construct a version of this theory with a doublet of vectors $I_\a{}^m$ parameterising breaking of the unimodularity constraint for the doublet of bi-vectors. Certainly, one would need to break 10d diffeormorphism symmetry explicitly according to the split $10=5+5$ similar to how KK reduction $11 \to 10$ allows to introduce $I^m$ in the first place.

\section*{Acknowledgments}

This work has been supported by Russian Science Foundation grant RSCF-20-72-10144 and in part by Russian Ministry of Education
and Science

\appendix

\section{Index conventions}
\label{app:index}

We are using the following conventions for indices

\begin{equation}
    \begin{aligned}
       &\hat{\mu}, \hat{\nu},\ldots = 1 \dots 10&& \mbox{ten directions, curved}; \\
       &\hat{a}, \hat{b},\ldots = 1 \dots 10&& \mbox{ten directions, flat}; \\       
       &\mu, \nu, \rho, \ldots = 1,\dots,5 && \mbox{external  directions (curved) of E6 ExFT}; \\
       &\bar{\mu}, \bar{\nu}, \bar{\rho}, \ldots 1,\dots,5 && \mbox{external   directions (flat) of E6 ExFT}; \\  
       &k, l, m, n,  \ldots = 1,\dots,5 && \mbox{internal directions (curved)}; \\       
       & a, b, c, d, \ldots = 1,\dots,5 && \mbox{internal directions (flat)}; \\  
       & M, N, K, L, \ldots = 1,\dots, 27 && \mbox{fundamental E6 (curved)};  \\
       & A, B, C, D, \ldots = 1,\dots, 27 && \mbox{fundamental E6 (flat)},\\
       & \a,\b,\dots = 1,2 && \mbox{fundamental SL(2) },\\
       & i_1,i_2, \dots = 1,\dots N, && \mbox{label }N\mbox{ Killing vectors}
    \end{aligned}
\end{equation}

For epsilon-tensors we adopt the following conventions
\begin{equation}
    \begin{aligned}
        \varepsilon^{\alpha \beta}\varepsilon_{\alpha \gamma}& =\delta_{\gamma}^{\beta}     , \\
        \varepsilon_{m n_1 k_1 l_1 p_1} \varepsilon^{m n_2 k_2 l_2 p_2}&=4! \delta^{ n_2 k_2 l_2 p_2}_{n_1 k_1 l_1 p_1} ,\\
        \varepsilon_{m l n k p}&=\sqrt{g}\,\epsilon_{m l n k p},
    \end{aligned}
\end{equation}
where
\begin{equation}
    \delta^{a_1\cdots a_p}_{b_1\cdots b_p} = \delta^{a_1}_{[b_1} \cdots \delta^{a_p}_{b_p]} 
 = \frac{1}{p!}\,\bigl(\delta^{a_1}_{b_1} \cdots \delta^{a_p}_{b_p}\pm \text{permutations}\bigr),
\end{equation}
and $g = \det ||g_{mn}||$ denotes determinant of what we call the internal metric. We also use $\sqrt{g} = e$ to condense notations.

\section{\texorpdfstring{$\rmE_{6(6)}$}{E6} algebra in the Type IIB decomposition}
\label{app:E6gens}

Generators 
 \begin{equation}
\begin{aligned}
 \big(T_{\a}{}^{k l}\big)_M{}^N& = {\begin{pmatrix}
 0 &   \delta_{m n}^{k l}\epsilon_{\a \beta} & 0 & 0 
\\
 0 &  0 & \frac{1}{2\sqrt{2}}\,\delta_\a{}^\b\epsilon^{m k l n_1 n_2} & 0 
\\
 0 & 0 & 0 & -\frac{1}{\sqrt{2}}\, \delta^{k l}_{ m_1 m_2}  \epsilon_{\a \beta} 
\\
 0 & 0 & 0 & 0 
\end{pmatrix}},\\
\big(T^{m_1 \ldots m_4}\big)_M{}^N &= \ {\begin{pmatrix}
 0 &  0 & \frac{1}{6\sqrt{2}} \epsilon^{n_1 n_2 [m_1 m_2 m_3} \delta_{m}{}^{m_4]} &  0 \\
 0 &  0 & 0 &  -\frac{1}{24} \delta_\beta{}^\alpha\, \epsilon^{m m_1 \ldots m_4} \\
 0 &  0 & 0 &  0 \\
 0 & 0 & 0 & 0
\end{pmatrix}} ,\\
 \big(T_{k l}{}^\delta\big)_M{}^N  &= {\begin{pmatrix}
 0 &  0 & 0 & 0 
\\
 - \delta^{m n}_{k l}\epsilon^{\d \alpha} &  0 & 0 & 0 
\\
 0 & \frac{1}{2\sqrt{2}}\delta_{\beta}{}^{\delta}\epsilon_{n kl m_1 m_2} & 0 & 0 
\\
 0 & 0 & -\frac{1}{\sqrt{2}}\, \delta^{n_1 n_2}_{k l}\epsilon^{\d \alpha} & 0 
 \end{pmatrix}},\\
 \big(T_{n_1 \ldots n_4}\big)_M{}^N&= \ {\begin{pmatrix}
 0 &  0 & 0 &  0 \\
 0 &  0 & 0 &  0\\
 \frac{1}{6\sqrt{2}} \epsilon_{m_1 m_2 [n_1 n_2 n_3} \delta_{ n_4]}{}^{n} &  0 & 0 &  0 \\
 0 & -\frac{1}{24} \delta_\beta{}^\alpha\, \epsilon_{n n_1 \ldots n_4} & 0 & 0
\end{pmatrix}} ,\\
\big(T_{k}{}^{l}\big)_M{}^N &=
 {\begin{pmatrix}
 \delta_{m}{}^{l} \delta_{k}{}^{n} +\frac{1}{3}\delta_k{}^{l}\delta_{m}{}^{n} & 0 & 0 & 0 \\
 0 & (\delta_{n}{}^{l} \delta_{k}{}^{m} +\frac{1}{3}\delta_k{}^{l}\delta_{n}{}^{m} )\delta_{\beta}{}^{\alpha} & 0 & 0 \\
 0 & 0 & -3\delta^{l n_1 n_2}_{k m_1 m_2} +\frac{1}{3}\delta_k{}^{l}\delta^{n_1 n_2}_{m_1 m_2} & 0 \\
 0 & 0 & 0 & -\frac{2}{3}\delta_k{}^{l} \delta_{\beta}{}^{\alpha}
\end{pmatrix}}, \\
\big(T_{\delta \gamma}\big)_M{}^N &=
 {\begin{pmatrix}
 0 & 0 & 0 & 0 \\
 0 & -\epsilon_{\beta (\delta } \delta_{ \gamma)}{}^{ \alpha} \delta_{n}{}^{m} & 0 & 0 \\
 0 & 0 & 0 & 0 \\
 0 & 0 & 0 & -\epsilon_{\beta(\delta} \delta_{ \gamma)}{}^{ \alpha} 
 \end{pmatrix}}.
 \end{aligned}
\end{equation}

\section{Transformation of the generalized fluxes}
\label{app:genfluxtransform}

To analyse transformations of the generalized flux components $Z_{AB}{}^C$ and $\Q_A$ under the 4-vector deformation \eqref{tens_def_4_rang} it is convenient to consider different orders in the deformation parameter $\r^{i_1i_2i_3i_4}$ separately. Denoting terms in a transformation of order $n$ by $\d_n$ we write
\begin{align}
\begin{aligned}
    \delta_1 Z_{A B}{}^{C}&=\\ \frac{\sqrt{5} }{48}e^{-3\phi} e \ d_{A B E} \Bigl\{&-24E_{k l}\,^{C} \partial_{m}{E_{n}\,^{E}}   k_{{i_{1}}}{}^{l} +24E_{k}\,^{C} \partial_{l}{E_{m n}\,^{E}}   k_{{i_{1}}}{}^{l} \\
    &  +48E_{k}\,^{(C} E_{l m}\,^{E)} \left( e^{-3\phi} e\right)^{-1}\partial_{n}{\left(e^{-3\phi} e\right)}  k_{{i_{1}}}{}^{l} \\
    &  + 4\sqrt{2} \epsilon_{\alpha \beta}  E^{q \beta E} E^{p \alpha C} \epsilon_{q m k l n} \partial_{p}{k_{{i_{1}}}{}^{l}} \\
    &  +2\sqrt{2} \left( e^{-3\phi} e\right)^{-1} \epsilon_{\alpha \beta} E^{q \beta (E} E^{|p \alpha| C)}  \partial_{p}{\left(e^{-3\phi} e\right)}   \epsilon_{q m k l n}  k_{{i_{1}}}{}^{l} \\
    &  - \sqrt{2}\epsilon_{\alpha \beta}E^{p \alpha C} \left(  \epsilon_{m n k l p} \partial_{q}{E^{q \beta E}} -\epsilon_{q m n k l} \partial_{p}{E^{q \beta E}}\right)   k_{{i_{1}}}{}^{l} \Bigr\}  K_{i_2 i_3 i_4}^{m \, n \, k} \rho^{{i_{1}} {i_{2}} {i_{3}} {i_{4}}}
\end{aligned}
\end{align}
\begin{equation}
\begin{aligned}
    \delta_2 Z_{A B}{}^{C}&= + \frac{3\sqrt{10}}{72}E_{k}\,^{C} E_{l}\,^{E} d_{ABE} \epsilon_{m n p q r}  \Big(-3 \partial_{t}{k_{{i_{1}}}{}^{m}}  e^{-3\phi} e k_{{i_{2}}}{}^{n}  k_{{i_{3}}}{}^{l}  \\
    & \quad \, - \partial_{t}{k_{{i_{1}}}{}^{l}} e^{-3\phi} e k_{{i_{2}}}{}^{m} k_{{i_{3}}}{}^{n}    -\partial_{t}{\left(e^{-3\phi} e\right)} k_{{i_{1}}}{}^{m} k_{{i_{2}}}{}^{n}  k_{{i_{3}}}{}^{l} \Big)K_{i_{4} i_5 i_6 i_7 i_8}^{p\,q\,r\,t\,k} \rho^{i_1{i_{2}} {i_{3}} {i_{4}} } \rho^{{i_{5}} {i_{6}} {i_{7}} {i_{8}}} \\
    & \quad \, + \frac{3\sqrt{10}}{72}E_{k}\,^{C}  \partial_{r}{E_{t}\,^{E}}  d_{A B E} e^{-3\phi} e \epsilon_{l m n p q} K_{i_1 \cdots i_8}^{l\,m\,n\,t\,p\,q\,r\,k} \rho^{{i_{1}} {i_{2}} {i_{3}} {i_{4}}} \rho^{{i_{5}} {i_{6}} {i_{7}} {i_{8}}} ,
\end{aligned}
\end{equation}
where we introduce a notation
\begin{equation}
    K_{i_1\dots i_n}^{m_1 \dots m_n} = k_{i_1}{}^{m_1}\dots k_{i_n}{}^{m_n}.
\end{equation}
The flux $\Q_A$ shifts only by terms linear in $\r$:\begin{equation}
\begin{aligned}
    \delta \Theta_{A}&= -\fr{3\sqrt{2}}{4}
    E^{m n}\,_{A} \epsilon_{m n k l p}\Bigg(3\partial_{q}{k_{{i_{1}}}{}^{k}} \rho^{{i_{1}} {i_{2}} {i_{3}} {i_{4}}}  k_{{i_{4}}}{}^{ q} + \partial_{q}{k_{{i_{4}}}{}^{ q}} \rho^{{i_{1}} {i_{2}} {i_{3}} {i_{4}}} k_{{i_{1}}}{}^{k}\Bigg)k_{{i_{2}}}{}^{l} k_{{i_{3}}}{}^{p}
    \\  &\quad \, - \frac{3\sqrt{2}}{4} \epsilon_{k l p m n}  \Big(\partial_{q}{E^{m n}\,_{A}}  - \frac{3}{5}E^{m n}\,_{A} e^{-1}\partial_{q}{e}   \Big)\rho^{{i_{1}} {i_{2}} {i_{3}} {i_{4}}}  K_{i_1 \cdots i_4}^{k \, l \, p \, q},
\end{aligned}
\end{equation}
where we used $-E_{M}{}^{B} \partial_{K} E^{M}{}_{B}=E^{-1} \partial_{K}{E} = - \frac{27}{5} e_{(5)}^{-1} \partial_{K}{e_{(5)}}$, $E=\sqrt{\det\left(\mM_{M N}\right)}$, $ e_{(5)}=\sqrt{\det g_{\mu \nu}}$. 

Given the condition that generalized Lie derivative \eqref{gen_lie} on the generalized vielbein $E_M{}^A$ along a vector $\L^M = (k^m,0,0,0)$ vanishes we arrive at the following identities, that will be heavily used in further calculations
\begin{equation}
\begin{aligned}
k_{{i}}{}\, ^{ l}\partial_{l}{E_{m n}^{A}}&= \ \ \, E_{m n}^{A}\partial_{l}{k_{{i}}{}\, ^{ l}}+E_{ n l}^{A}\partial_{m}{k_{{i}}{}\, ^{ l}}+E_{ l m}^{A}\partial_{n}{k_{{i}}{}\, ^{ l}}\\
k_{{i}}{}\, ^{ l}\partial_{l}{E^{m n}_{A}}&= -\left(E^{m n}_{A}\partial_{l}{k_{{i}}{}\, ^{ l}}+E^{ n l}_{A}\partial_{l}{k_{i}\,^{m}}+E^{  l m}_{A}\partial_{l}{k_{i}{}\,^{n}}\right)\\
k_{i}{}\,^{q}\partial_{q}{\left(e^{-3\phi} e\right)} &=-e^{-3\phi} e\partial_{q}{k_{i}{}\,^{q}}\\
k_{i}\,^{m}\partial_{m}{E^{n}_{A}}&= \ \ \, E^{m}_{A}\partial_{m}{k_{i}{}\,^{n}}\\
k_{i}\,^{m}\partial_{m}{E_{n}^{A}}&=-E_{m}^{A}\partial_{n}{k_{i}{}\,^{m}}\\
\partial_{q}{e_{(5)}} k_{i}{}\,^{q}&=- \frac{5}{3}e_{(5)}\partial_{q}{k_{i}{}\,^{q}}\\
k_{i}{}\,^{n}\partial_{n}{E_{ m \beta A}}&=-E_{ n \beta A}\partial_{m}{k_{i}{}\,^{n}}\\
k_{i}{}\,^{n}\partial_{n}{E^{ m \beta A}}&=  \ \ \, E^{ n \beta A}\partial_{n}{k_{i}{}\,^{m}}\\
k_{i}{}\,^{n}\partial_{n}{E^{\alpha E}}&= \ \ \, \partial_{n}{k_{i}{}\,^{n}}E^{\alpha E}.
\end{aligned}
\label{all_weights}\end{equation}
Using \eqref{all_weights} we obtain
\begin{equation}
\begin{aligned}
    \delta_1 Z_{A B}{}^{C}&={} 
    - \sqrt{5}\left(2E_{k}\,^{(E} E_{l m}\,^{C)}-E_{l}\,^{E} E_{k m}\,^{C}\right)d_{A B E} e^{-3\phi} e  \partial_{n}{k_{{i_{1}}}{}^{l}} \rho^{ {i_{1}} {i_{2}} {i_{3}} {i_{4}} } K_{i_2 i_3 i_4}^{k \, m \, n}
    %\\ &\quad \ +\sqrt{5} E_{l}\,^{E} E_{k m}\,^{C} d_{A B E} e^{-3\phi} e \partial_{n}{k_{{i_{1}}}{}^{l}} \rho^{{i_{1}} {i_{2}} {i_{3}} {i_{4}}}   k_{{i_{2}}}{}^{k} k_{{i_{3}}}{}^{m} k_{{i_{4}}}{}^{n}
    %\\ &\quad \ -\sqrt{5} E_{k}\,^{C} E_{l m}\,^{E} \partial_{n}{k_{{i_{1}}}{}^{l}} \rho^{ {i_{1}} {i_{2}} {i_{3}} {i_{4}}} d_{A B E} e^{-3\phi} e k_{{i_{2}}}{}^{k} k_{{i_{3}}}{}^{m} k_{{i_{4}}}{}^{n} 
    \\ &\quad \ +\frac{\sqrt{10}}{12}\epsilon_{\alpha \beta} E^{k \alpha C} E^{l \beta E}  d_{A B E} e^{-3\phi} e\left( \epsilon_{l m n p q} \partial_{k}{k_{{i_{1}}}{}^{m}}-\epsilon_{m n p q k} \partial_{l}{k_{{i_{1}}}{}^{m}}+\epsilon_{ n p q k l} \partial_{m}{k_{{i_{1}}}{}^{m}}\right) \rho^{{i_{1}} {i_{2}} {i_{3}} {i_{4}}}  K_{i_2 i_3 i_4}^{n \, p \, q}
    \end{aligned}
    \end{equation}
    \begin{equation}
    \begin{aligned}
    \delta_2 Z_{A B}{}^{C}= \frac{3\sqrt{10}}{72} E_{k}\,^{C} E_{l}\,^{E}d_{A B E} e^{-3\phi} e\epsilon_{m n p q r}\Big(& 3   k_{{i_{1}}}{}^{l} k_{{i_{6}}}{}^{t}\partial_{t}{k_{{i_{2}}}{}^{m}}+  k_{{i_{2}}}{}^{m}\left(k_{{i_{6}}}{}^{t} \partial_{t}{k_{{i_{1}}}{}^{l}}-k_{{i_{1}}}{}^{t} \partial_{t}{k_{{i_{6}}}{}^{l}}\right) 
    \\ &+  \partial_{t}{k_{{i_{6}}}{}^{t}}  k_{{i_{2}}}{}^{m} k_{{i_{1}}}{}^{l} \Big) \rho^{{i_{1}} {i_{2}} {i_{3}} {i_{4}}} \rho^{{i_{5}} {i_{6}} {i_{7}} {i_{8}}} K_{i_3 i_4 i_5 i_7 i_8}^{n \, p \, k \, q \, r}
    %k_{{i_{3}}}{}^{n} k_{{i_{4}}}^{p} k_{{i_{5}}}^{k}  k_{{i_{7}}}^{q} k_{{i_{8}}}^{r} 
\end{aligned}
\end{equation}
\begin{equation}    
\begin{aligned}
    \delta \Theta_{A}={} \frac{3\sqrt{2}}{4}E^{k l}\,_{A}\Big(&- 3  \epsilon_{k l m n p} \partial_{q}{k_{{i_{1}}}{}^{m}}  + 2  \epsilon_{k m n p q} \partial_{l}{k_{{i_{1}}}{}^{m}} -  \epsilon_{k l n p q} \partial_{m}{k_{{i_{1}}}{}^{m}}   \Big)\rho^{{i_{1}} {i_{2}} {i_{3}} {i_{4}}} K_{i_2 i_3 i_4}^{n \, p \, q}
\end{aligned}
\end{equation}
Using the Killing vector algebra
\begin{align}
    k_{[i_1}{}^{m}\partial_{m}{k_{i_2]}{}^{n}}=\frac{1}{2} f_{i_1 i_2}{}^{i_3}k_{i_3}{}^{n}
\end{align}
and some antisymmetrizations in six indices we arrive at the final form of the transformations
\begin{equation}
\begin{aligned}
\delta Z_{A B}^{C}
    &=\frac{ \sqrt{10} }{24}{e^{-3\phi} e}d_{A B E} E_{k}\,^{C} E_{l}\,^{E} \epsilon_{m n p q r}\left( 3\rho^{{i_{1}} {i_{2}} {i_{3}} {i_{4}}} \rho^{{i_{5}} {i_{6}} {i_{7}} {i_{8}}}   f_{{i_{1}} {i_{7}}}\,^{{i_{9}}}-\right. 
    \\&-\left. \rho^{{i_{1}} {i_{2}} {i_{3}} {i_{4}}} \rho^{{i_{5}} {i_{6}} {i_{7}} {i_{9}}}   f_{{i_{1}} {i_{7}}}\,^{{i_{8}}}\right) 
    K_{i_2 \cdots i_6 i_8 i_9}^{m \, n\, k \,p \,q \, l \, r}\\
    {}
    &- \frac{\sqrt{5}}{2}\left(E_{k}\,^{(E} E_{m n}\,^{C)}-E_{n}\,^{E} E_{m k}\,^{C}\right) d_{A B E} e^{-3\phi} e  \rho^{{i_{1}} {i_{2}} {i_{3}} {i_{4}}} f_{{i_{1}} {i_{2}}}\,^{{i_{5}}}  K_{i_3 i_4 i_5}^{m\, k\, n}
    %\\&+\frac{\sqrt{5}}{2}E_{k}\,^{E} E_{m n}\,^{C} \rho^{{i_{1}} {i_{2}} {i_{3}} {i_{4}}} d_{A B E} e^{-3\phi} e f_{{i_{1}} {i_{2}}}\,^{{i_{5}}} k_{{i_{3}}}{}^{m} k_{{i_{4}}}{}^{n} k_{{i_{5}}}\,^{k} 
    %\\&- \frac{\sqrt{5}}{2}E_{k}\,^{C} E_{m n}\,^{E} \rho^{{i_{1}} {i_{2}} {i_{3}} {i_{4}}} d_{A B E} e^{-3\phi} e f_{{i_{1}} {i_{2}}}\,^{{i_{5}}} k_{{i_{3}}}{}^{m} k_{{i_{4}}}^{k} k_{{i_{5}}}\,^{n} 
    \\&+\frac{\sqrt{10}}{8}E^{k \alpha C} E^{l \beta E} \epsilon_{\alpha \beta} d_{A B E} e^{-3\phi} e \epsilon_{k l m n p} \rho^{{i_{1}} {i_{2}} {i_{3}} {i_{4}}}  f_{{i_{1}} {i_{2}}}\,^{{i_{5}}} K_{i_3 i_4 i_5}^{m \,n \,p}
    \\&+\frac{\sqrt{10}}{24}E_{k}\,^{C} E_{l}\,^{E} d_{A B E} e^{-3\phi} e \epsilon_{m n p q r}  \rho^{{i_{5}} {i_{6}} {i_{7}} {i_{8}}} \rho^{{i_{1}} {i_{2}} {i_{3}} {i_{4}}} f_{{i_{1}} {i_{2}}}\,^{{i_{9}}} K_{i_3 \cdots i_9}^{m \,k\, n \,p \,q \,l \,r}
    \end{aligned}  
\end{equation}
\begin{equation}
    \delta \Theta_{A}= {}\frac{9\sqrt{2}}{4}E^{k l}\,_{A} \epsilon_{k l m n p} \rho^{{i_{1}} {i_{2}} {i_{3}} {i_{4}}} f_{{i_{1}} {i_{4}}}\,^{{i_{5}}}   K_{i_2 i_3 i_5}^{m \, n \, p}
\end{equation}

As one can see, enough condition of equality to zero deformations of components of $\f$ flux may be written as system of algebraic conditions:
\begin{align}
    3\rho^{{i_{1}} {i_{4}} [{i_{2}} {i_{3}} } \rho^{{i_{5}} {i_{6}} |{i_{7}} {i_{8}}|}   f_{{i_{1}} {i_{7}}}\,^{{i_{9}}]}+ \rho^{{i_{1}} {i_{4}} [{i_{2}    } {i_{3}} } \rho^{{i_{5}} {i_{6}}  {i_{9}}] {i_{7}}}   f_{{i_{1}} {i_{7}}}\,^{{i_{8}}}=0, \label{four_gcybe_app}\\
    \rho^{ {i_{1}}  {i_{4}} {i_{2}} [{i_{3}}} f_{{i_{1}} {i_{4}}}\,^{{i_{5}}]}=0
    \label{four_unimod_app}
\end{align}

\bibliography{bib.bib}

\bibliographystyle{utphys.bst}

\end{document}